\documentclass[12pt]{article}

\usepackage{a4}
\usepackage{subeqnarray}
\usepackage[dvips]{epsfig}
\usepackage{epic}
\usepackage{natbib}
\newcommand{\be}{\begin{equation}}
\newcommand{\ee}{\end{equation}}
\newcommand{\bae}{\begin{eqnarray}}
\newcommand{\eae}{\end{eqnarray}}
\newcommand{\bse}{\begin{subeqnarray}}
\newcommand{\ese}{\end{subeqnarray}}

\begin{document}


\centerline{\Large\bf  The  sensitivity and stability  of the
ocean's} \centerline{\Large\bf  thermohaline circulation to finite
amplitude  perturbations}

\vspace{0.5cm} \centerline{\sc  Mu Mu } \centerline{\it \small
LASG, Institute of Atmospheric Physics, Chinese Academy of
Sciences, Beijing 100029, China} \vspace{0.25cm} \centerline{\sc
Liang SUN} \centerline{\it \small Department of Mechanics
Engineering,} \centerline{\it \small University of Science and
Technology of China, Hefei 230027, China; } \centerline{and}
\centerline{\it \small LASG, Institute of Atmospheric Physics,
Chinese Academy of Sciences, Beijing 100029, China}
\vspace{0.25cm} \centerline{and} \vspace{0.25cm} \centerline{\sc
Henk A. Dijkstra} \centerline{\it \small Institute for Marine and
Atmospheric research Utrecht, Utrecht University, the Netherlands}
\centerline{and} \centerline{\it \small Department of Atmospheric
Science, Colorado State University, Fort Collins, CO, USA}

\newpage
\begin{abstract}
Within a simple model context, the  sensitivity and stability of
the thermohaline circulation to finite amplitude  perturbations is
studied. A  new approach is used  to  tackle this nonlinear
problem. The method is based on the computation of the so-called
Conditional Nonlinear Optimal Perturbation (CNOP) which is a
nonlinear generalization of the linear singular vector approach
(LSV). It is shown  that linearly stable thermohaline circulation
states can become nonlinearly unstable and  the properties of the
perturbations with optimal nonlinear growth are determined. An
asymmetric nonlinear response to perturbations exists with respect
to the sign of finite amplitude freshwater perturbations,  on both
thermally dominated and salinity  dominated thermohaline flows.
This asymmetry is due to the nonlinear interaction of the
perturbations through advective processes.
\end{abstract}
\vspace{2cm} {\bf Key words:} thermohaline circulation,
conditional optimal nonlinear perturbation,  nonlinear stability,
sensitivity

\newpage

\section{Introduction}

A recurrent theme in fundamental research on climate variability
is the sensitivity  of the ocean's thermohaline circulation. When
state-of-the-art climate models are used to calculate projections
of future climate states  as a response to different emission
scenarios of greenhouse gases, a substantial  spread in the model
results is found.  One of the reasons of this spread is the
diverse behavior of the  thermohaline circulation
\cite[]{TAR2001}.

The sensitivity of the thermohaline circulation is caused by
several  feedbacks induced by  the physical processes
that determine  the evolution of  the thermohaline flow. One of
these feedbacks is the salt-advection feedback which is caused by
the fact that salt is transported by the thermohaline flow, but in
turn influences the density difference which drives this flow. The
salt-advection feedback can be conceptually understood in a
two-box model \cite[]{Stommel1961} where it is shown to cause
multiple equilibria and hysteresis behavior.

In many models of the global ocean circulation, it appears that
several equilibrium states may exist under similar forcing
conditions. When the present equilibrium state, with about 16 Sv
Atlantic overturning,   is subjected to a quasi-steady freshwater
input in the North Atlantic, eventually the circulation may
collapse. In this collapsed state, there is deepwater formation in
the Southern Ocean instead of in the North Atlantic and the
formation of North Atlantic Deep Water (NADW) has ceased
\cite[]{Stocker1992, Rahmstorf1995b, Manabe1999}. As this multiple
equilibrium regime seems to be present in many ocean models, it is
important to determine whether transitions between the different
states can occur due to finite amplitude perturbations.

In a variant of the Stommel-model for which the temperature
relaxation is fast, \cite{Cessi1994} studied the transition
behavior between the different equilibria. In this case, there are
only three equilibrium states, of which one unstable. In the
deterministic case, she finds that a finite amplitude perturbation
of the freshwater flux can shift the system into an alternate
state and that the minimum amplitude depends on the duration of
the disturbance. Regardless of the duration, however,   the
amplitude of the disturbance has to exceed  a certain value
for a transition to occur. Under stochastic white-noise forcing,
there are occasional transitions from one equilibrium to another
as well as fluctuations around each state.

In \cite{Timmermann2000}, the effect of multiplicative noise
(through fast fluctuations in the meridional thermal temperature
gradient)  on the variability in  a box model similar to that in
\cite{Cessi1994} has been studied. It was found that the stability
properties of the thermohaline circulation depend on the noise
level.  Red noise can introduce new equilibria that do
not have a  deterministic counterpart.

Another line of studies
uses box models that show intrinsic variability  because of the
existence of an oscillatory mode in the eigenspectrum  of the
linear operator. \cite{Griffies1995} show that noise is able to
excite an otherwise damped eigenmode of the thermohaline
circulation.   \cite{Tziperman2002} study the non-normal growth of
perturbations on the thermally driven state
and identify two physical mechanisms associated  with the
transient amplification of these perturbations.

Stochastic noise can have a significant effect on the mean states
of the thermohaline circulation and their stability
\cite[]{Hasselmann1976,  Palmer2001,  Velez-Belchi2001,
Tziperman2002}. Some of these mechanisms are intrinsically linear,
such as the effects of non-normal growth considered in
\cite{Tziperman2002}. Others are   essentially  nonlinear
mechanisms,  such as those causing the noise-induced transitions
reported in \cite{Timmermann2000}.

To study linear amplification mechanisms, the  linear singular
vector (LSV) method is often used, with main applications  to
predictability studies \cite[]{Xue1997a,Xue1997b, Thompson1998}.
\cite{Knutti2002}, for example, found that the sensitivity of the ocean
circulation to perturbations severely limits the predictability of
the future thermohaline circulation when approaching the
bifurcation point. The LSV approach, however,  cannot provide
critical boundaries on finite amplitude stability of the
thermohaline ocean circulation.

In a system which potentially has multiple equilibria and internal
oscillatory modes, its response to a finite amplitude perturbation
on a particular steady state is a difficult nonlinear problem. In
this paper,  we determine   the  nonlinear stability  boundaries
of  linearly stable thermohaline flow states within a simple box
model of the thermohaline circulation. To compute these
boundaries, we use  the concept of  the Conditional Nonlinear
Optimal Perturbation (CNOP)  and study optimal {\it nonlinear}
growth over a certain given time $\tau$. We extend results on
linear optimal growth properties of  perturbations on both
thermally and salinity dominated thermohaline flows to the
nonlinear case. We find that  there is an  asymmetric nonlinear
response of these flows with respect to the sign of the finite
amplitude freshwater perturbation and  describe a physical
mechanism that explains this asymmetry.

\section{Model and methodology }

\subsection*{\it a. Model}

To illustrate the approach, the theory is applied to a 2-box model
of the thermohaline circulation \cite[]{Stommel1961}.  This  model
consists of an equatorial box  and a polar box which contain well
mixed water of different temperatures and salinities due to an
equatorial-to-pole gradient in atmospheric surface forcing. Flow
between the boxes is assumed proportional to the density
difference between the boxes and, with a linear equation of state,
related to the temperature and salinity differences in the boxes.

When the balances of heat and salt are nondimensionalized, the
governing dimensionless equations  (we use the notation in
chapter 3 of \cite{DijkstraB2000})  can be written as
\bse
\frac{d T}{dt} &=& \eta_1 - T ( 1 + \mid T - S  \mid ) \\
\frac{d S}{dt} &=& \eta_2  - S (\eta_3 + \mid T - S  \mid )
\label{e:2box_d}
\ese
where $T = T_e - T_p, ~S = S_e - S_p$ are the dimensionless
temperature and salinity  difference between the equatorial and
polar box and  $\Psi = T - S$ is the dimensionless flow rate.
Three parameters appear in the equations (\ref{e:2box_d}): the
parameter $\eta_1$ measures the strength of the thermal forcing,
$\eta_2$ that of the freshwater forcing and $\eta_3$ is the ratio
of the relaxation times of temperature and salinity to the surface
forcing. Steady states of the equations (\ref{e:2box_d}) are
indicated with a temperature of $\bar{T}$, a salinity of $\bar{S}$
and a flow rate $\bar{\Psi} = \bar{T} - \bar{S}$. A steady state
is called  thermally-dominated when $\bar{\Psi}>0$, i.e.  a
negative  equatorial-to-pole temperature gradient exists
dominating  the density.  A steady state is called
salinity-dominated when $\bar{\Psi}<0$,  i.e.  a negative
equatorial-to-pole salinity  gradient exists dominating  the
density.

It is  well known that the equations (\ref{e:2box_d}) have
multiple steady states for certain parameter values. Here, we fix
$\eta_1 = 3.0$, $\eta_3 = 0.2$ and use $\eta_2$ as control
parameter. The bifurcation
diagram for these parameter values is shown in Fig.~\ref{f:1} as a
plot of $\bar{\Psi}$ versus $\eta_2$. Solid curves indicate
linearly stable steady states, whereas the states on the dashed
curve are unstable. There are thermally-driven (hereafter TH)
stable steady states ($\bar{\Psi} > 0$) and salinity-driven
(hereafter SA, ie., the circulation is salinity-dominated) stable
steady states ($\bar{\Psi}<0$). The saddle-node bifurcation points
occur at $\eta_2 = 0.600$ and $\eta_2 = 1.052$, and bound the
interval in $\eta_2$ where multiple equilibria occur. Suppose that
this bifurcation diagram represents both the present overturning
state (on the stable branch with $\bar{\Psi} > 0$) and  the
collapsed state (on the stable branch with $\bar{\Psi} < 0$). To
study  the nonlinear transition behavior of the thermohaline
flows from the TH state to the SA state and vice versa, we consider the
evolution of finite amplitude perturbations on the stable states.

The nonlinear equation governing the evolution of perturbations
can be derived from equation (\ref{e:2box_d}).  If the steady
state ($\bar{T}, \bar{S}$) is given and  $T'=T-\bar{T}$,
$S'=S-\bar{S}$ are the perturbations of temperature and salinity,
then it is found that
\bse
\frac{dT'}{dt} &=& -(2|\bar{\Psi}|+1)T'+sign(\bar{\Psi})[\bar{T}\,S'-\bar{S}T'-T'(T'-S')] \\
\frac{dS'}{dt} &=&-(2|\bar{\Psi}|+\eta_3)S'+sign(\bar{\Psi})[\bar{T}\,S'-\bar{S}T'-S'(T'-S')]
\label{e:thpetur1}
\ese
where $sign(\bar{\Psi})$ is sign of steady flow rate $\bar{\Psi}$.
If the perturbations are sufficiently small, such  that the
nonlinear part of the equations (\ref{e:thpetur1}) can be
neglected, we find the tangent linear equation  governing the
evolution of small perturbations as
  \bse
    \frac{dT'}{dt} &=& -(2|\bar{\Psi}|+1)T'+sign(\bar{\Psi})(\bar{T}\,S'-\bar{S}T') \\
     \frac{dS'}{dt}
     &=&-(2|\bar{\Psi}|+\eta_3)S'+sign(\bar{\Psi})(\bar{T}\,S'-\bar{S}T')
\label{e:thpetur2} \ese

\subsection*{\it b. Conditional Nonlinear Optimal Perturbation}

To study nonlinear mechanisms of amplification,
\cite{Barkmeijer1996} modified the LSV technique and \cite{Mu2000}
proposed the concept of nonlinear singular vectors (NSVs) and
nonlinear singular values (NSVAs). These concepts were
successfully applied by \cite{Mu2001} and \cite{Durbiano2001}  to
study finite amplitude stability of flows in  two-dimensional
quasi-geostrophic and shallow-water models, respectively. In
\cite{Mu2003a}, the concept of the conditional nonlinear optimal
perturbation (CNOP) was introduced and applied to study the
``spring predictability barrier" in the El Nino-Southern
Oscillation (ENSO), using a  simple equatorial ocean-atmosphere
model. The ``spring predictability barrier"  refers to the
dramatical decline of the prediction skills for most of the ENSO
models during the Northern Hemisphere  (NH) springtime. The CNOP
can also be employed to estimate the prediction errors of an El
Ni\~no or a La Ni\~na event \cite[]{Mu2003b}.

As readers may not be familiar with this concept, we give a brief
introduction to CNOP. Considering the nonlinear evolution of
initial perturbations governed by (\ref{e:thpetur1}). In general,
assume that the equations governing the evolution of
perturbations can be written  as:
\be\begin{array}{ll} \left\{
\begin{array}{l}
\displaystyle\frac{\partial {\textbf{\textit{x}}}}{\partial t} +
F(\textbf{\textit{x}};\bar{\textbf{\textit{x}}}) = 0,
\\[0.2cm]
 {\textbf{\textit{x}}}|_{t=0} = {\textbf{\textit{x}}}_0, \\
\end{array}
\right. & \ \ \ {\rm in}\ \ \ \Omega\times [0,t_e]
\end{array}
\label{e:cnopmodel}\ee
where $t$ is time, ${\textbf{\textit{x}}}(t)=(x_1(t), x_2(t), ...,
x_n(t))$ is the perturbation state vector and $F$ is  a nonlinear
differentiable operator. Furthermore, ${\textbf{\textit{x}}_0}$ is
the initial perturbation, $\bar{\textbf{\textit{x}}}$ is the basic
state, $({\textbf{\textit{x}}},t)\in \Omega\times [0,t_e]$ with
$\Omega$ a domain in $R^n$, and $t_e< + \infty$.

Suppose the initial value problem (\ref{e:cnopmodel}) is well-posed
and the nonlinear propagator $M$ is defined as the evolution
operator  of (\ref{e:cnopmodel}) which determines a trajectory from
the initial time $t = 0$  to time $t_e$. Hence, for fixed $t_e>0$, the solution
${\textbf{\textit{x}}}(t_e)=M({\textbf{\textit{x}}_0;\bar{\textbf{\textit{x}}}})(t_e)$
is well-defined, i.e.
\be
 \textbf{\textit{x}}(t_e) = M(\textbf{\textit{x}}_0;\bar{\textbf{\textit{x}}})(t_e)
\label{e:cnopsol}\ee
So $\textbf{\textit{x}}(t_e)$ describes the
evolution of the initial perturbation ${\textbf{\textit{x}}}_0$.

For a chosen norm $\|\cdot\|$ measuring ${\textbf{\textit{x}}}$,
the perturbation ${\textbf{\textit{x}}}^{\*}_{0\delta}$ is called
the Conditional Nonlinear Optimal Perturbation (CNOP) with constraint
condition $\|{\textbf{\textit{x}}}_0\|\leq\delta$ , if and only if
\be J({\textbf{\textit{x}}}^{\*}_{0\delta}) =
\max_{\|{\textbf{\textit{x}}}_0\| \leq
\delta}J({\textbf{\textit{x}}}_0)\ee where \be
J({\textbf{\textit{x}}}_0)=\|M({\textbf{\textit{x}}}_0;\bar{\textbf{\textit{x}}})(t_e)\|\ee

The CNOP is the initial perturbation whose nonlinear evolution
attains the maximal value of the functional $J$ at time $t_e$ with
the constraint conditions; in this sense we call it ``optimal".
The CNOP can be regarded as the most (nonlinearly) unstable initial
perturbation superposed on the basic state. With the same
constraint conditions, the larger the nonlinear evolution of the
CNOP is, the  more unstable the basic state is.
In general,  it is  difficult to obtain an analytical
expression of the CNOP. Instead we look for the  numerical solution,
by  solving a constraint nonlinear optimization problem.

To calculate the CNOP  the norm
\be \|{\textbf{\textit{x}}}_0\| =
\sqrt{(T'_{0})^2+(S'_{0})^2}\label{e:norm}\ee
is used. Using a fourth-order Runge-Kutta scheme with a time step
$dt=0.001$, perturbation solutions $(T',S')$ are obtained
numerically by integrating the model (\ref{e:thpetur1}) up to a
time $t_e$ and the magnitude of the perturbation is calculated.
Next, the Sequential Quadratic Programming (SQP) method is applied
to obtain the CNOP numerically; the SQP method   is briefly
described in the Appendix.

To compare the CNOPs with the LSVs, the latter are also computed
using the theory of linear singular vector analysis \cite[]{ChenYQ1997}.
We also use the  norm (\ref{e:norm}) in this analysis and first  solve
(\ref{e:thpetur2}) to obtain the linear evolution of initial
perturbations. Subsequently,  the singular vector decomposition (SVD)
is used to determine the linear singular vectors (LSVs) of the model.

\section{Stability and sensitivity analysis }

In this section, we compute  the CNOPs to study the sensitivity of
the thermohaline circulation to finite amplitude freshwater
perturbations in the two-box model.  Two problems are studied: (i)
the nonlinear development of the finite amplitude perturbations
for fixed parameters in the model and (ii)  the nonlinear
stability of the steady states as parameters are changed. Both
thermally-dominated steady states  ($\bar{\Psi}>0$)  and
salinity-dominated ones ($\bar{\Psi}<0$) are investigated. The
initial perturbation $\textbf{\textit{x}}_0$ is written as
$\textbf{\textit{x}}_0 =(T'_{0},S'_{0}) =
(\delta\cos\theta,\delta\sin\theta)$ , where $\delta$ is magnitude
of initial perturbation and  $\theta$ the angle of the initial
vector with the x-axis.

\subsection{Finite amplitude evolution of the TH state}

For the  thermally-driven stable steady state, we consider the
state $\bar{T}=1.875,\bar{S}=1.275,\bar{\Psi}=0.6$ (shown as point
"A" in Fig.~\ref{f:1})  with the fixed parameters $\eta_1=3.0$,
$\eta_2=1.02$, $\eta_3=0.2$.    We choose  $t_e=2.5$ and use
$\delta = 0.3$ as a maximum norm (in the norm (\ref{e:norm}))  of
the perturbations.  The time $t_e$ is about half the time  the solution
takes to equilibrate to steady state from a particular initial
perturbation. The amplitude $\delta = 0.3$ is about $10\%$ of the
typical amplitude of the steady state  of temperature and salinity
$(\bar{T},\bar{S})$. For $\theta$ in the range
$\pi/4<\theta<5\pi/4$, the initial perturbation flow has
$\Psi'(0)<0$. As this is typically caused by a freshwater flux perturbation
in the polar box, we refer to the perturbation as being of  freshwater type.
For other angles $\theta \in [0,2\pi]$ , the initial perturbation flow has
$\Psi'(0)>0$, which is typically caused by a salt flux perturbation
in the polar box and we refer to it as being of  salinity type.

Using equation (\ref{e:thpetur1}) and (\ref{e:thpetur2}), both
CNOPs  and LSVs are computed versus the constraint condition
$\delta$, respectively. The numerical results, plotted
in Fig.~\ref{f:2},  indicate  that the
CNOPs are located at the circle $\|\textbf{\textit{x}}_0\|=\delta$,
which is the boundary of ball
$\|\textbf{\textit{x}}_0\|\leq\delta$.  The directions of the  LSVs, which are
independent of $\delta$, have constant values of $\theta_1=1.948$
(dashed line) and $\theta_2=5.089$ (not shown). The value of $\theta$ for
the CNOPs (solid curve) increases monotonically over the $\delta$ interval
0.01 to 0.3. The difference between CNOPs and LSVs is
relatively small when $\delta$ is small.

Integrating the model (\ref{e:thpetur1}) with CNOPs and LSVs as
initial conditions, respectively. we obtain their evolutions at
time $t_e$, which are denoted as "CNOP-N" and "LSV-N"; these
are shown in Fig.~\ref{f:3}. For comparison, the linear
evolution  of the LSVs are also obtained by integrating the model
(\ref{e:thpetur2}) with the LSV as an initial condition; this is
denoted as "LSV-L" in Fig.~\ref{f:3}. It is clear that the evolution
of the CNOPs is nonlinear in $\delta$, while  "LSV-L" only increases
linearly. The line of "LSV-N" is between "CNOP-N" and "LSV-L", but the
difference between "LSV-N" and "CNOP-N" is hardly distinguishable
over the whole  $\delta$ interval. Though this difference is not significant in this TH
state, it is significant in the following investigation for SA
state. In fact, it is very hard to know this without previous
calculation.

Note  that since our numerical results demonstrate that the CNOPs
are all located on the boundary $ \|{\textbf{\textit{x}}}_0\|
=\delta$,  we are able to show the sensitivity of THC to finite
amplitude perturbations of specific fixed amplitude $\delta=0.2$.
In Fig.~\ref{f:4} the value of $J$ at $t_e = 2.5$  is shown for
the linear and the nonlinear evolutions of the initial
perturbations obtained by (\ref{e:thpetur1}) and
(\ref{e:thpetur2}). For the linear case (dashed line in
Fig.~\ref{f:4}), there are two optimal linear initial
perturbations $\theta_1=1.948$ and $\theta_2=5.089$ with the same
value of $J$, $J(\delta,\theta_1)=J(\delta,\theta_2) =0.16484$,
which are the LSVs. Note that $\theta_2-\theta_1=\pi$, which means
that in the linear case perturbations with $\Psi' > 0$  and $\Psi'
< 0$ behave similarly (and hence symmetrically with respect to the
sign of $\Psi'$). For the nonlinear model (solid curve in
Fig.~\ref{f:4}), there is one global optimal nonlinear initial
perturbation with $\theta_3=1.979$, which is the CNOP, and one
local optimal nonlinear initial perturbations at $\theta_4 =
5.058$, with values of $J(\delta,\theta_3)=0.22413$ and
$J(\delta,\theta_4)=0.13052$, respectively. The results in
Fig.~\ref{f:4} for $\delta = 0.2$ coincide with the results in
Fig.~\ref{f:2}.

There is another difference between the linear and nonlinear
evolution of the perturbations. When the initial perturbations are
freshwater ($\Psi' < 0$), the nonlinear evolution leads to a
larger amplitude than the linear evolution. When the initial
perturbations are saline ($\Psi' > 0$), the nonlinear evolution
leads to a smaller amplitude than the linear  evolution. For
example, the initial perturbations with $\theta_1$ and $\theta_3$
are such that $\Psi' < 0$, while the initial perturbations with
$\theta_2$ and $\theta_4$ have $\Psi'>0$.

The values of $J/\delta$ obtained by integrating  (\ref{e:thpetur1})
with the CNOPs as initial condition are shown for different
$\delta$ in Fig.~\ref{f:5}a.  The corresponding evolution of
$\Psi$ is plotted in   Fig.~\ref{f:5}b. To relate the result in
Fig. 5a to previous ones, consider the value of $J/\delta$ at $t =
2.5$ on the curve of $\delta = 0.2$. In Fig. 4, the maximum of $J$
is 0.224 and hence $J/\delta = 0.224/0.2 = 1.12$. It follows from
the Figs.~\ref{f:5} that for the CNOP with $\delta=0.01$, the flow rate
$\Psi$ recovers to the steady state $\bar{\Psi}=0.6$ rapidly. For the
CNOP with a larger initial amplitude ($\delta=0.1, 0.2$), it takes much
longer for the thermohaline circulation to recover to steady
state. This is different  from a linear analysis where
the evolution is the same for all optimal initial perturbations
\cite[]{Tziperman2002}.

 In summary, the results for the TH state show
that fresh perturbations, with $\Psi'<0$, are more amplified
though nonlinear mechanisms than saline perturbations, with
$\Psi'>0$. This is consistent with the notion  that perturbations
which  move the system towards a bifurcation point will be more
amplified through non-linear mechanisms than perturbations that
move the system away from a bifurcation point.

\subsection{Finite amplitude stability of the SA state}

We consider the salinity-dominated SA state ($\bar{T}= 2.674,
\bar{S}=2.796, \bar{\Psi}=-0.122$) for a slightly smaller value of
$\eta_2$ than for the thermally-dominated TH state in the previous
section ($\eta_1=3.0$, $\eta_2=0.9$, $\eta_3=0.2$). It is
indicated as point "B" in Fig.~\ref{f:1}.  Again for the time $t_e
= 2.5$, using the corresponding equations (\ref{e:thpetur1}) and
(\ref{e:thpetur2}), both CNOPs and LSVs are computed versus the
constraint condition $\delta$, respectively and the results are
plotted in Fig.~\ref{f:6}.

Similar to the results in Fig.~\ref{f:2}, the directions of the
LSVs, which are independent of $\delta$, have constant values of
$\theta_1=2.796$ and $\theta_2=5.938$ (dashed line). The line for
$\theta_1$ is similar to $\theta_2$, and is not drawn in
Fig.~\ref{f:6}. The direction of CNOPs (solid curve) increase
monotonously with $\delta$ varying from 0.01 to 0.125. Then,
$\theta$ drops down to $2.857$ at $\delta=0.125$ and increase
slightly with $\delta$ in the interval $0.125<\delta<0.17$. After
that, $\theta$ jumps up to $5.195$ at $\delta=0.17$ and increases
slightly with $\delta$ in $0.17<\delta<0.22$.

Using equations (\ref{e:thpetur1}) and (\ref{e:thpetur2}), the
evolutions of both the CNOPs and LSVs of the SA state are shown in
Fig.~\ref{f:7}. All the three kinds of evolutions are
approximately the same when the initial perturbations are
relatively small. But for larger $\delta$ the difference between
"LSV-N" and "CNOP-N" is remarkably larger than that of the TH
state (Fig.~\ref{f:3}). The values of $J$  of "CNOP-N" are always
larger  than those of both "LSV-L" and "LSV-N" for $\delta >
0.17$.

The values of $J$  at a time $t_e=2.5$ for all $\theta$ show
(Fig.~\ref{f:8}a)  two  optimal linear initial perturbations
$\theta_1=2.796$ and $\theta_2=5.938$ with
$J(\delta,\theta_1)=J(\delta,\theta_2)=0.0526$. Again, because
$\theta_2-\theta_1=\pi$ there is the symmetry in response with
respect to the sign of $\Psi'$. For nonlinear evolutions, there is
one globally  optimal  initial perturbation at $\theta_3=5.246$
(the CNOP)  and two locally optimal  initial perturbations at
$\theta_4=0.251$ and $\theta_5=2.890$, with J-values of
$J(\delta,\theta_3)=0.0963$, $J(\delta,\theta_4)=0.0432$ and
$J(\delta,\theta_5)=0.0503$, respectively. The initial
perturbations with  $\theta_1$ and $\theta_5$ are of freshwater
type ($\Psi'<0$), while the initial perturbations of $\theta_2$,
$\theta_3$ and $\theta_4$ are of salinity type ($\Psi'>0$).

To understand the difference between the maxima located at
$\theta_3$, $\theta_4$ and $\theta_5$, a contour graph of
$J(\theta,\delta)$ is drawn in Fig.~\ref{f:8}b. It is clear from
Fig.~\ref{f:8}b that there are three groups of local maxima which
are indicated  by the dashed lines. When $\delta<0.125$ there are
only two local maxima and the CNOP is located in the regime
$5\pi/4<\theta<2\pi$. When $0.125<\delta<0.17$, the CNOP jumps
from the interval $5\pi/4<\theta<2\pi$ to the interval
$3\pi/4<\theta<5\pi/4$. This coincides with the jumping behavior
of $\theta$ in Fig.~\ref{f:6}. When  $\delta>0.17$, there are
three local maxima and the CNOP jumps from the interval
$3\pi/4<\theta<5\pi/4$ to a new interval $5<\theta<6$. Both  jumps
are also shown in Fig.~\ref{f:6} and $J$ has a remarkable increase
after the second jump (Fig.~\ref{f:7}).

Also for the SA state, the value of $J/\delta$ along trajectories
obtained by integrating equations (\ref{e:thpetur1}) using the
CNOPs as initial conditions  are shown for different $\delta$ in
Fig.~\ref{f:9}a.  The corresponding evolution of $\Psi$ is plotted
in  Fig.~\ref{f:9}b. For $t = 2.5$ and $\delta = 0.2$, the value
of $J/\delta = 0.0963/0.2 = 0.48$, where 0.0963 is the maximum in
Fig.~\ref{f:8}a. The flow rate $\Psi$ recovers to the steady state
(whose value is $-0.122$) shortly after being disturbed with a
small amplitude CNOP ($\delta=0.01$). It takes much longer for the
thermohaline circulation to recover to the steady state after
being disturbed with a larger amplitude ($\delta=0.1,0.2$),
respectively. The larger the CNOP is, the larger the transient
effect. In contrast to the TH state, there is now an oscillatory
attraction to the SA state, already described in
\cite{Stommel1961}.

In both the salinity-dominated SA state and the
thermally-dominated TH state, the CNOP always moves the system
towards the bifurcation point. The SA state (TH state) has an
asymmetry in the nonlinear amplification of disturbances, with a
larger amplification for those with $\Psi'>0$ ($\Psi'<0$).

\subsection{Sensitivity along the bifurcation diagram}

\begin{figure}[htpb]
\begin{minipage}[t]{\linewidth}
\centering\epsfig{file=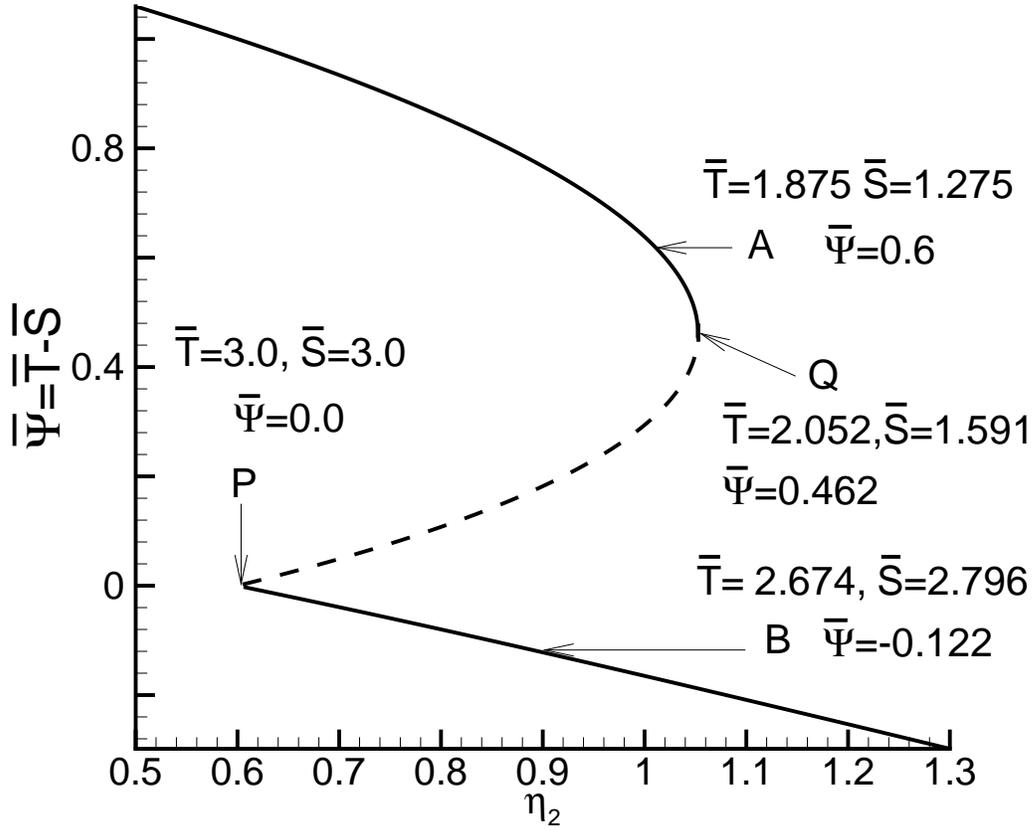, width=\linewidth}
\end{minipage}\hfill
\caption{\em \small The bifurcation diagram of the Stommel two-box
model for $\eta_1 = 3.0$ and $\eta_3 = 0.2$.  The points labelled
A and B represent the thermally-driven steady state  and
salinity-driven steady state, respectively, considered in section
3.  The points labelled P and Q represent the bifurcation points
of the model, respectively. Solid curves indicate linearly stable
steady states, whereas the states on the dashed curve are
unstable. There are thermally-driven (TH) stable steady states
($\bar{\Psi} > 0$) and salinity-driven (SA, ie., the circulation
is salinity-dominated) stable steady states ($\bar{\Psi}<0$).}
\label{f:1}
\end{figure}

\begin{figure}[htpb]
\begin{minipage}[t]{\linewidth}
\centering\epsfig{file=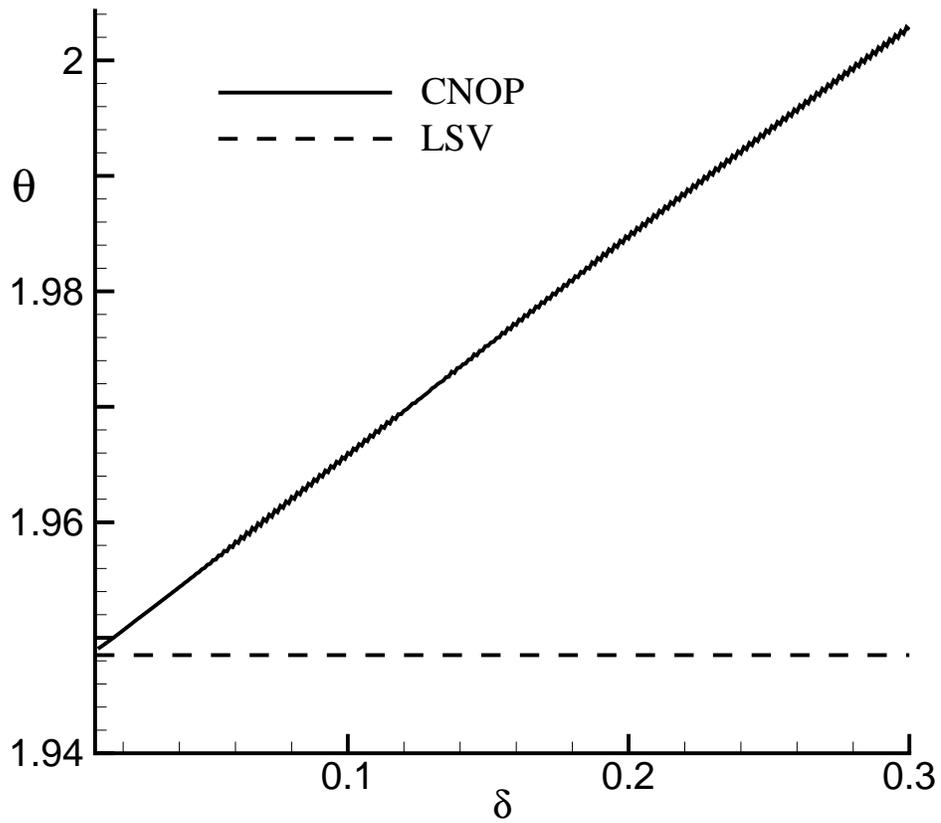, width=\linewidth}
\end{minipage}\hfill
\caption{\em \small The values of $\theta$ for both the linear
singular vectors (LSV, dashed line) and for the Conditional
Nonlinear Optimal Perturbation (CNOP, solid line) of the
thermally-driven state under the conditions $\delta \leq 0.3$ and
$t_e = 2.5$.  } \label{f:2}
\end{figure}

\begin{figure}[htpb]
\begin{minipage}[t]{\linewidth}
\centering\epsfig{file=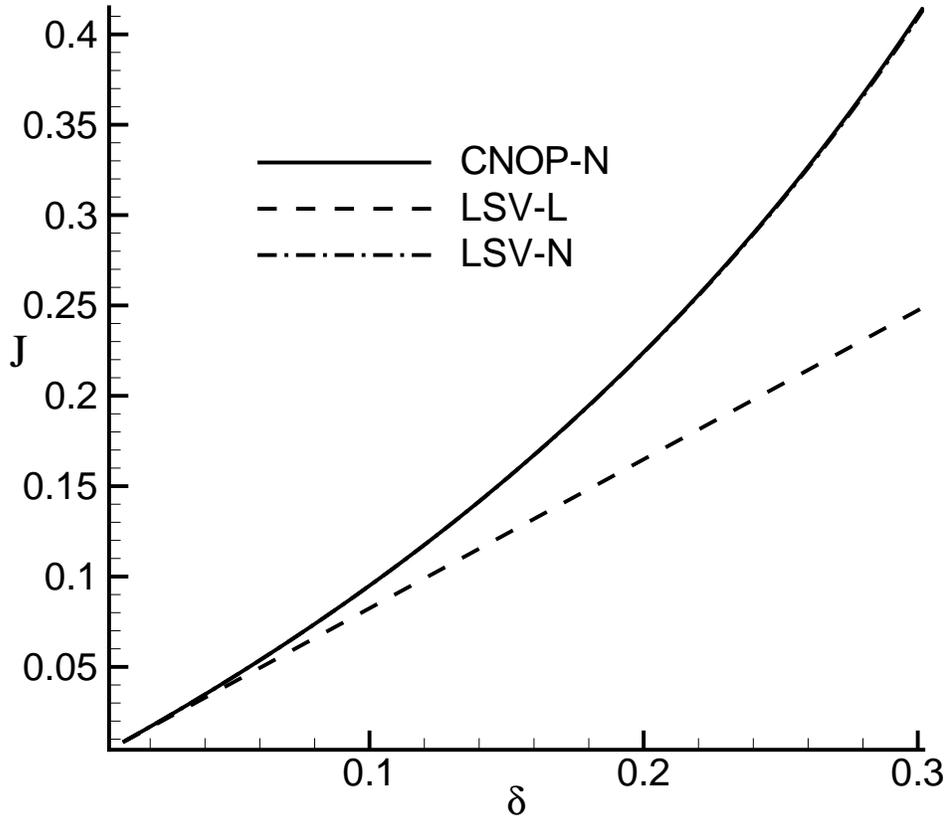, width=\linewidth}
\end{minipage}\hfill
\caption{\em \small  Values of $J$ at  the endpoints  of
trajectories  at time $t_e=2.5$ for different values of $\delta$.
These trajectories started either with Conditional Nonlinear
Optimal Perturbation (CNOP-N, solid curve) and linear singular
vectors (LSV-N,dash-dotted curve, hardly distinguishable from the
solid curve) perturbing the thermally-driven state. Also included
are the endpoints when the tangent linear model is integrated with
the linear singular vectors as initial perturbation (LSV-L, dashed
curve ). } \label{f:3}
\end{figure}

Even if  a TH or SA state is linearly stable, it can become
nonlinearly unstable due to finite amplitude perturbations. The
methodology of CNOP provides a means to assess the nonlinear
stability thresholds of the thermohaline flows; here this is
shown for the two-box model (2).  Thereto, we compute the CNOPs under
different  $\delta$ constraints for linearly stable TH and SA
states  along the bifurcation  diagram in Fig.~\ref{f:1}.

Along the TH branch, we  vary $\eta_2$  from 1.043 to 1.046 with
step 0.001 and thereby approach the saddle-node bifurcation
(Fig.~\ref{f:1}).  For each value of $\eta_2$, the CNOPs  are
obtained under the constraint that the magnitude of initial
perturbations is less 0.2 ($\delta = 0.2$) and again  the time
$t_e=2.5$. The trajectories of the  CNOPs are calculated by
integrating the model (\ref{e:thpetur1}) (Fig.~\ref{f:10}a). Next,
the corresponding flow rates $\Psi$ are drawn in Fig.~\ref{f:10}b.
Both figures indicate that the CNOPs damp after a while for the
steady states labelled $A_1$, $A_2$ and $A_3$. While these three
states are consequently nonlinearly stable, the CNOP for  steady
state $A_4$ ($\eta_2=1.046$) increases in time, which implies that
this steady state is nonlinearly unstable (although it is linearly
stable) to perturbations with $\delta=0.2$.

  From the above results, it follows that for each value of $\eta_2$,
in the multiple equilibria regime of Fig.~\ref{f:1}, a critical
value of $\delta$, say  $\delta_c$, must exists such that the TH
state is nonlinearly unstable.   $\delta_c$ is defined as the
smallest magnitude of a finite amplitude perturbation which
induces a transition from the TH state to the SA state. The larger
the value of $\delta_c$, the more stable the steady state is.
Using the CNOP method, the values of $\delta_c$ can be computed
and the results for the TH states (from $\eta_2 = 0.95$ up to  the
saddle-node bifurcation at $\eta_2 = 1.052$ ) are shown in
Fig.~\ref{f:11}. The curve separates the plane into two parts. For
the regime under the curve the steady state is nonlinearly stable
and for the regime above the curve it is nonlinearly unstable.
When the bifurcation point $Q$ in Fig.~\ref{f:1} is approached,
$\delta_c$ decrease more and more quickly. The critical value
$\delta_c$ reduces to zero sharply, as  $\eta_2$ approaches the
bifurcation point. This explains how the steady states lose their
stability when the  bifurcation point $Q$ is reached.

The same calculations are performed for steady states on the SA
branch, when $\eta_2$ varies from 0.75 to the value 0.60 at the
saddle-node bifurcation. The CNOPs are obtained under the
constraint that the magnitude of initial perturbations is less
than 0.1 ($\delta=0.1$) with time $t_e=2.5$. The trajectories of
the CNOPs at each steady state (of which $J$ and $\Psi$ are
plotted in Fig.~\ref{f:12}) indicate that the CNOPs damp for the
steady states labelled $B_1$, $B_2$ and $B_3$. Although there is
oscillatory behavior, these states are nonlinearly stable.
However,  the  evolution of CNOP for steady state $B_4$
($\eta_2=0.70$) increases, which implies that the  SA state is
nonlinearly unstable to this finite amplitude perturbation
(although it is linearly stable).

The critical boundaries $\delta_c$ for the SA states
(Fig.~\ref{f:13}) show that  $\delta_c$ decreases monotonically
with $\eta_2$ and becomes  zero at saddle-node bifurcation
($\eta_2 = 0.6$). Similar to Fig.~\ref{f:11}, the curve separates
the plane into two parts. For the regime under the curve the
steady state is nonlinearly stable and for the regime above the
curve it is nonlinearly unstable. When $\eta_2$ decreases from
$1.0$ to $0.6$, the SA steady states approach the  bifurcation
point $P$ in Fig.~\ref{f:1}, and the critical value $\delta_c$
tends to zero. This explains how the steady states lose their
stability at the  bifurcation point $P$.

\section{Summary and Discussion}

Within a simple two-box model, we have addressed the sensitivity
and nonlinear stability of (linearly stable) steady states of the
thermohaline circulation.  One of the remarkable results obtained
by the CNOP approach is the asymmetry in the nonlinear
amplification of perturbations with $\Psi'<0$ (interpreted as a
freshwater perturbation in the northern North Atlantic) and $\Psi'
>0$ (interpreted as a salt  perturbation in the northern North
Atlantic).

\begin{figure}[htpb]
\begin{minipage}[t]{\linewidth}
\centering\epsfig{file=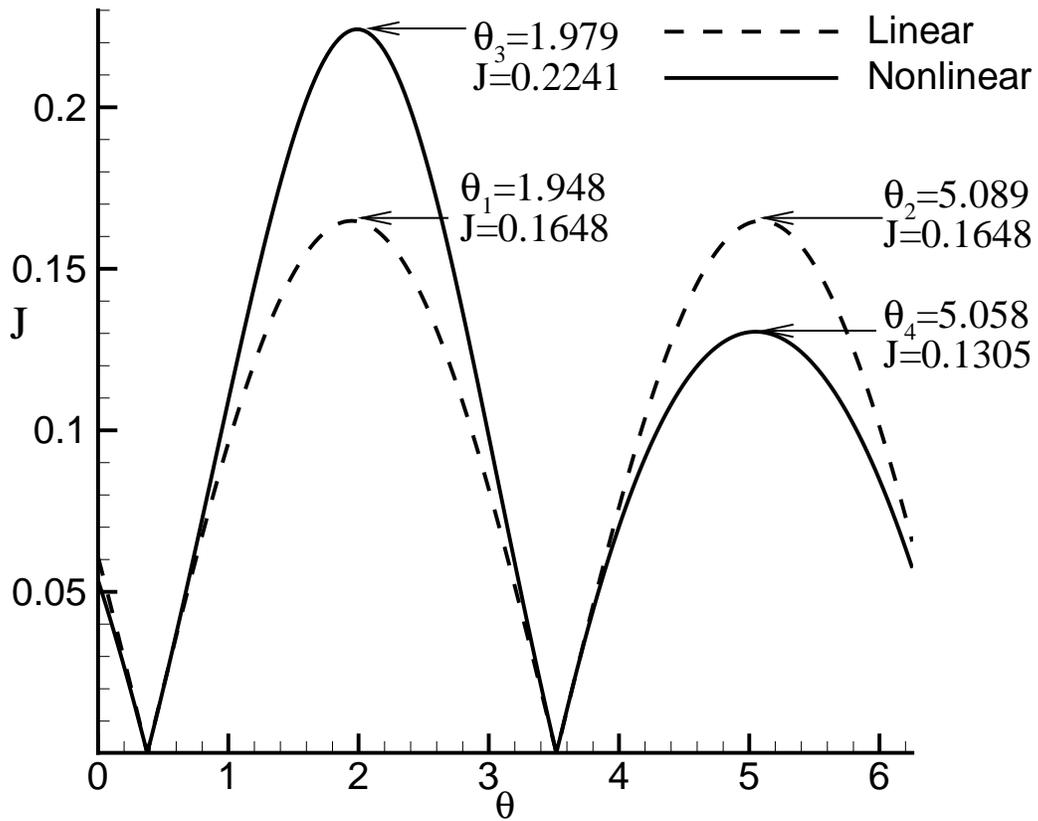, width=\linewidth}
\end{minipage}\hfill
\caption{\em \small The magnitude of perturbations $J$ obtained at
$t_e = 2.5$ for the the evolutions of perturbations of the
thermally-driven state in the tangent linear model and nonlinear
model.  The initial perturbations have the form $(T'(0), S'(0)) =
(\delta\cos\theta,\delta\sin\theta)$ with  $\delta=0.2$. }
\label{f:4}
\end{figure}

\begin{figure}[htpb]
\begin{minipage}[t]{\linewidth}
\centering\epsfig{file=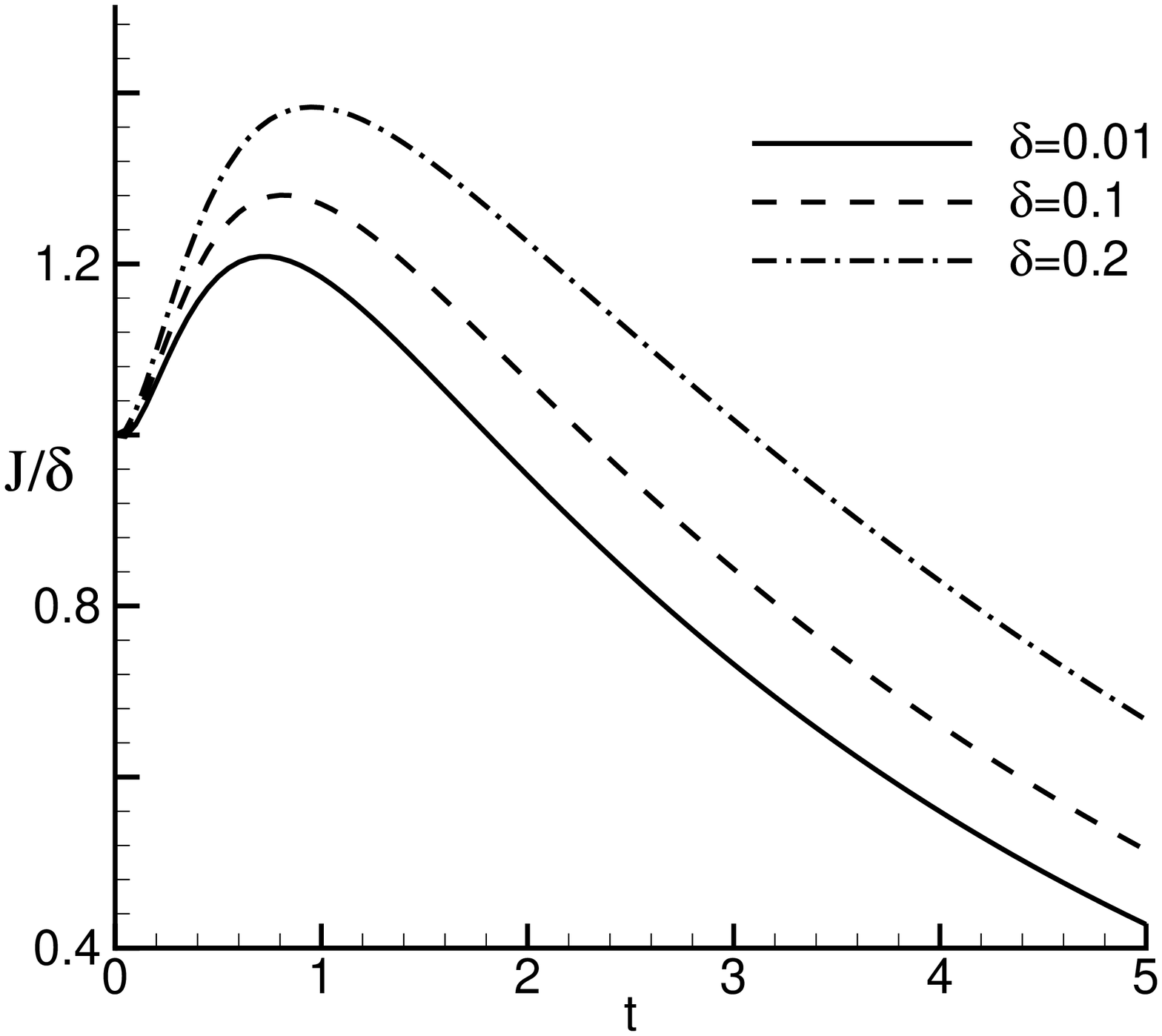,width=0.65\linewidth}  (a) \\
\end{minipage}\hfill
\begin{minipage}[t]{\linewidth}
\centering\epsfig{file=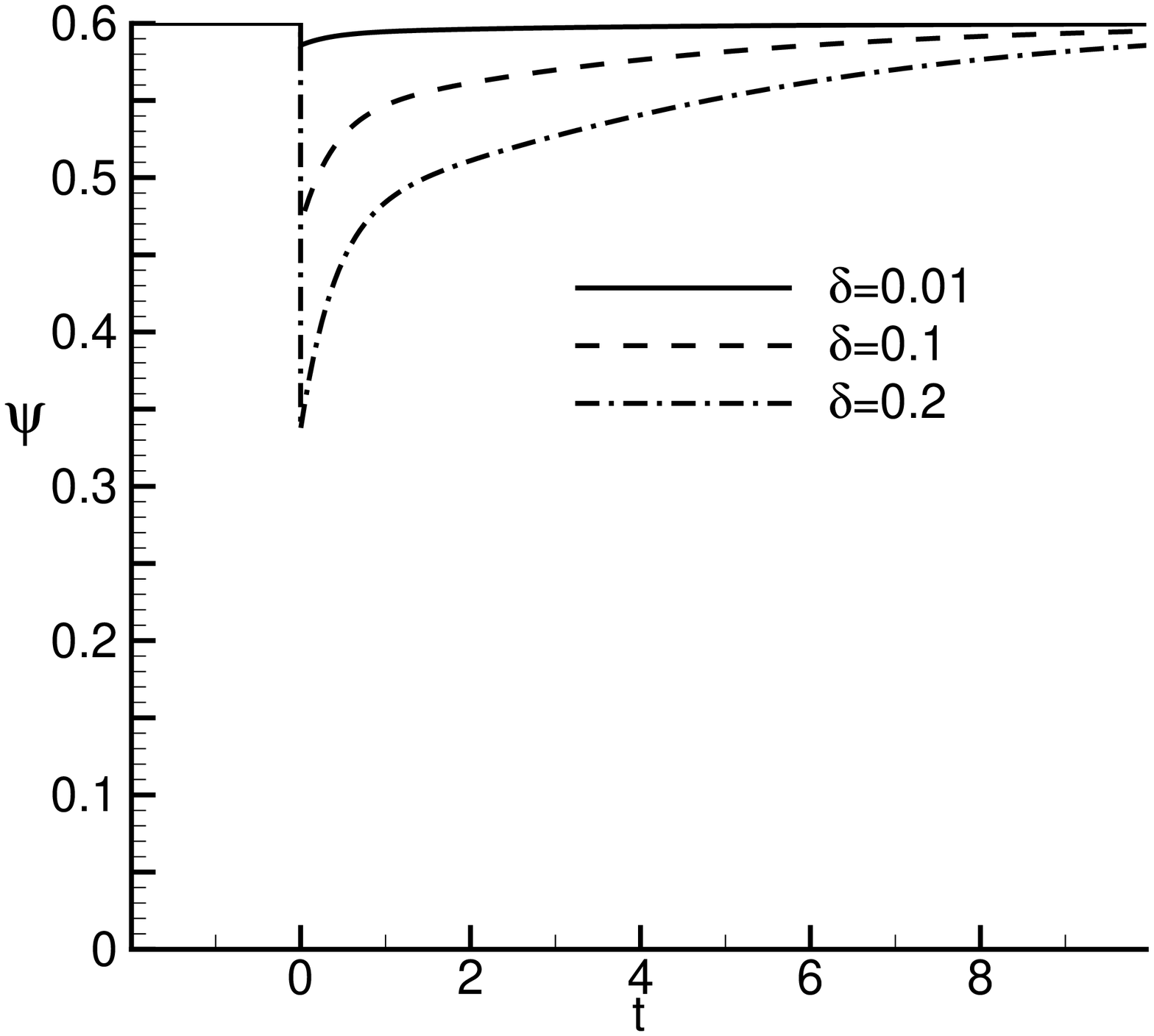,width=0.65\linewidth}  (b)
\end{minipage}\hfill
\caption{\em \small  Values of (a) perturbation growth $J/\delta$
and (b) flow stream function $\Psi$ along the trajectories
computed with Conditional Nonlinear Optimal Perturbation (CNOP,
solid curve) initial conditions superposed on the thermally-driven
state. The different curves are for different values of $\delta$.
} \label{f:5}
\end{figure}

\begin{figure}[htpb]
\begin{minipage}[t]{\linewidth}
\centering\epsfig{file=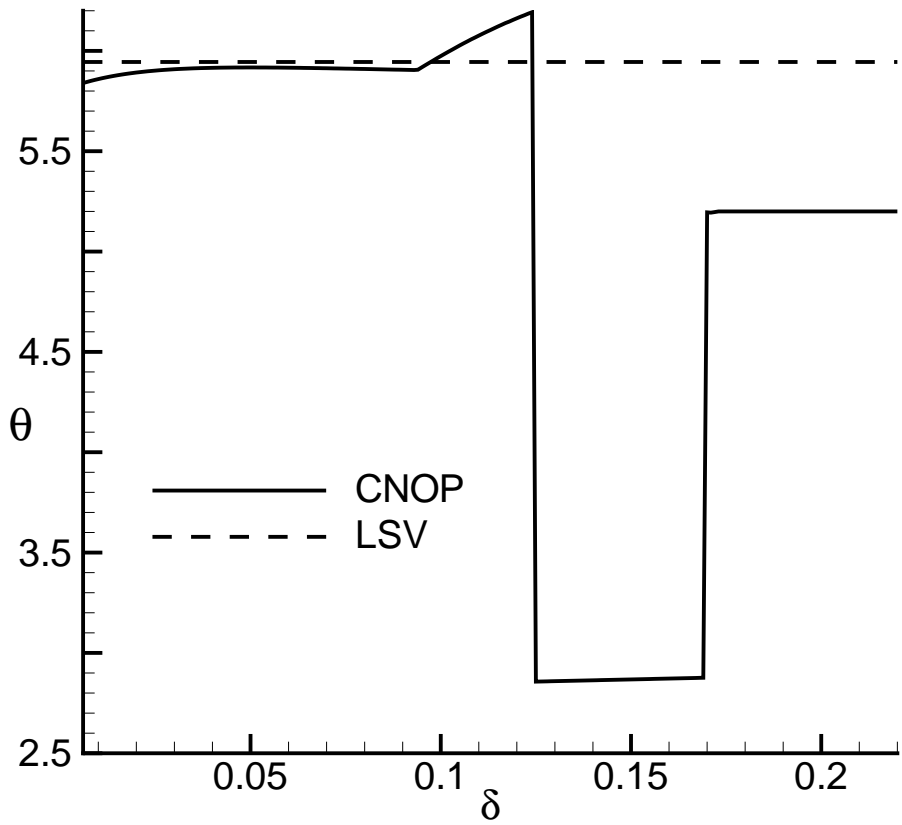, width=\linewidth}
\end{minipage}\hfill
\caption{\em \small The values of $\theta$ for both the linear
singular vectors (LSV, dashed line) and for the Conditional
Nonlinear Optimal Perturbations (CNOP) of the salinity-driven
state under the conditions $\delta \leq 0.22$ and $t_e = 2.5$ . }
\label{f:6}
\end{figure}

\begin{figure}[htpb]
\begin{minipage}[t]{\linewidth}
\centering\epsfig{file=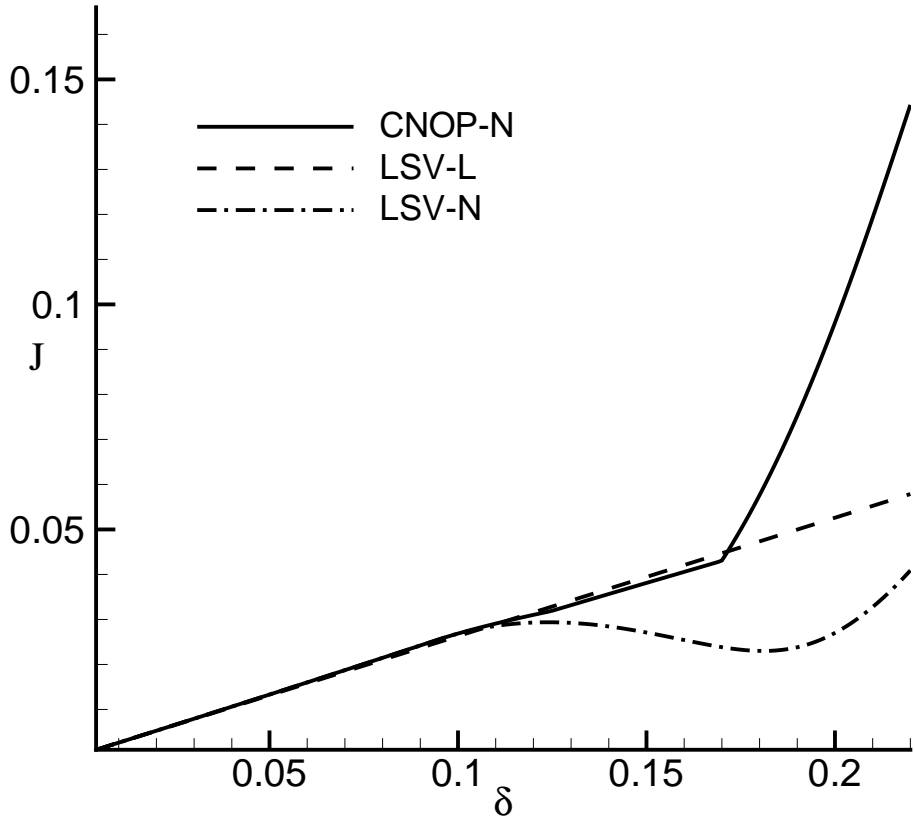,width=\linewidth}
\end{minipage}\hfill
\caption{\em \small The magnitude of perturbation $J$ at  the
endpoints of trajectories at time $t_e=2.5$ for different values
of $\delta$. These trajectories started either with Conditional
Nonlinear Optimal Perturbation (CNOP-N, solid curve) and linear
singular vectors (LSV-N, dash-dotted line) perturbing the
salinity-driven state. Also included are the endpoints when the
tangent linear model is integrated with the linear singular
vectors as initial perturbation (LSV-L, dashed line). }
\label{f:7}
\end{figure}

\begin{figure}[htpb]
\begin{minipage}[t]{\linewidth}
\centering\epsfig{file=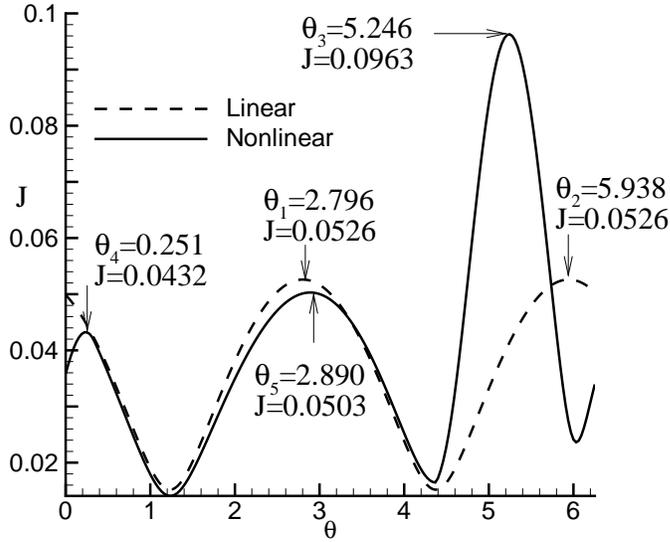,width=0.65\linewidth}  (a) \\
\end{minipage}\hfill
\begin{minipage}[t]{\linewidth}
\centering\epsfig{file=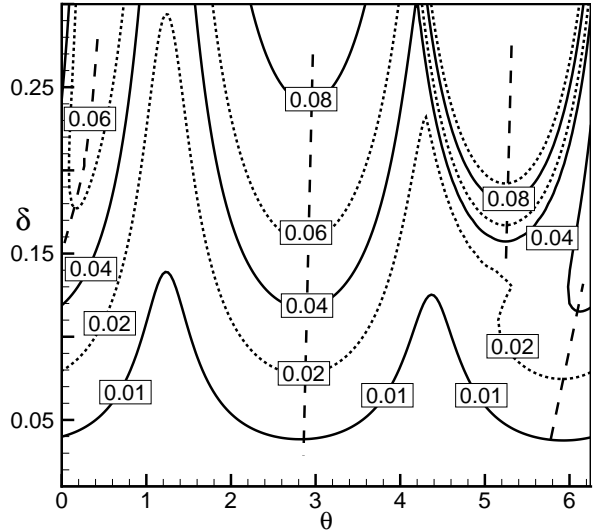,width=0.65\linewidth}  (b)
\end{minipage}\hfill
\caption{\em \small (a) Values of perturbation magnitude $J$
obtained at $t_e = 2.5$ for the the evolutions of perturbations of
the salinity-driven state in the tangent linear model and
nonlinear model. The initial perturbations have the form $(T'(0),
S'(0)) = (\delta\cos\theta,\delta\sin\theta)$ with  $\delta=0.2$ .
 And (b) the contour plot of $J(\theta,\delta)$, with contour
interval of 0.02 from $J=0.02$ to $J=0.08$ (solid curve and dotted
curve). The three group local maxima are indicated as the  dashed
curves .} \label{f:8}
\end{figure}

\begin{figure}[htpb]
\begin{minipage}[t]{\linewidth}
\centering\epsfig{file=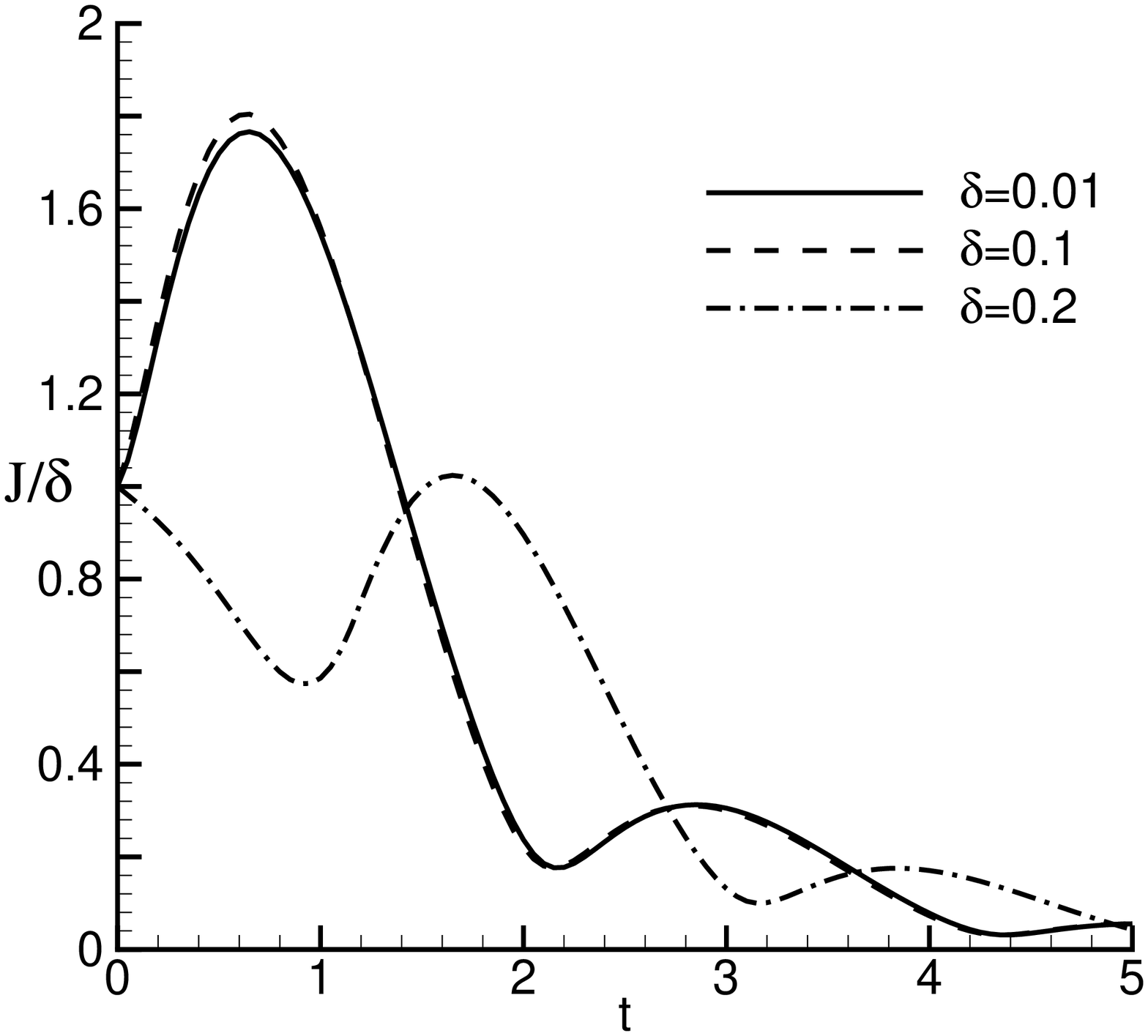,width=0.65\linewidth}  (a) \\
\end{minipage}\hfill
\begin{minipage}[t]{\linewidth}
\centering\epsfig{file=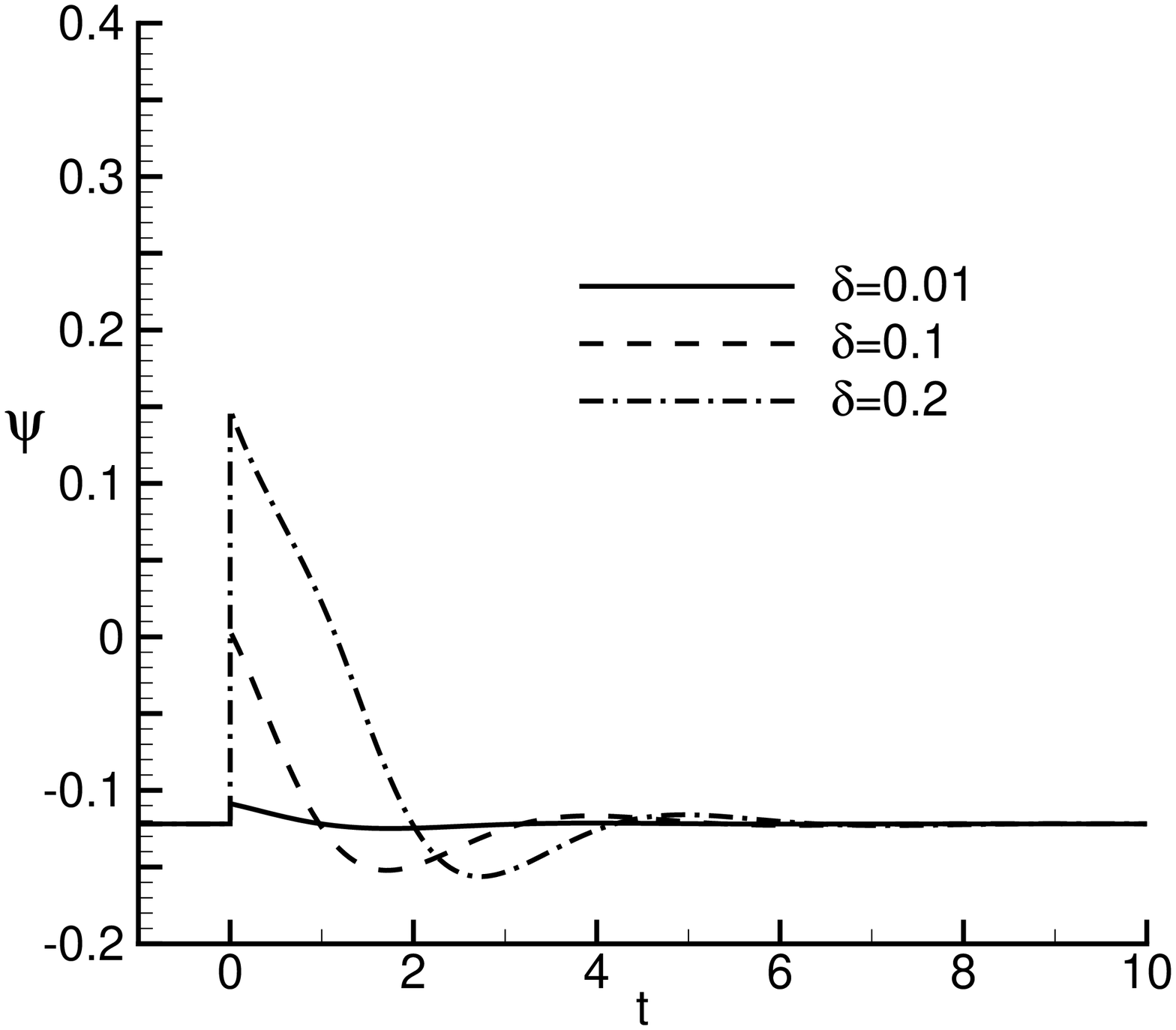,width=0.65\linewidth}  (b)
\end{minipage}\hfill
\caption{\em \small   Values of (a) perturbation growth $J/\delta$
and (b) flow stream function $\Psi$ along the trajectories
computed with Conditional Nonlinear Optimal Perturbation (CNOP),
initial conditions superposed on the salinity-driven state. The
different curves are for different values of $\delta$.   }
\label{f:9}
\end{figure}

\begin{figure}[htpb]
\begin{minipage}[t]{\linewidth}
\centering\epsfig{file=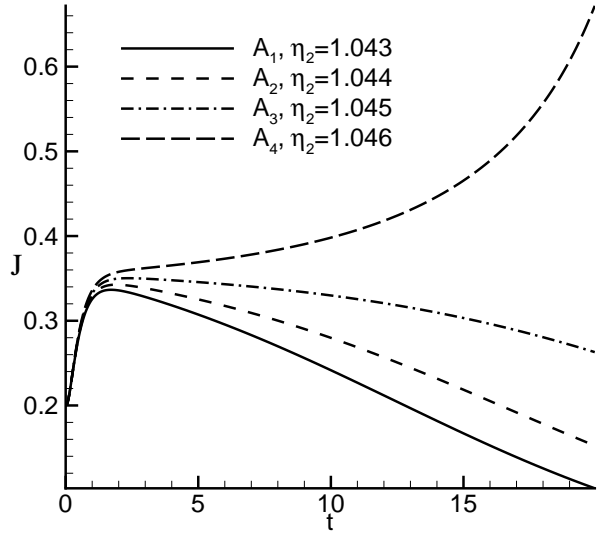,width=0.65\linewidth}  (a) \\
\end{minipage}\hfill
\begin{minipage}[t]{\linewidth}
\centering\epsfig{file=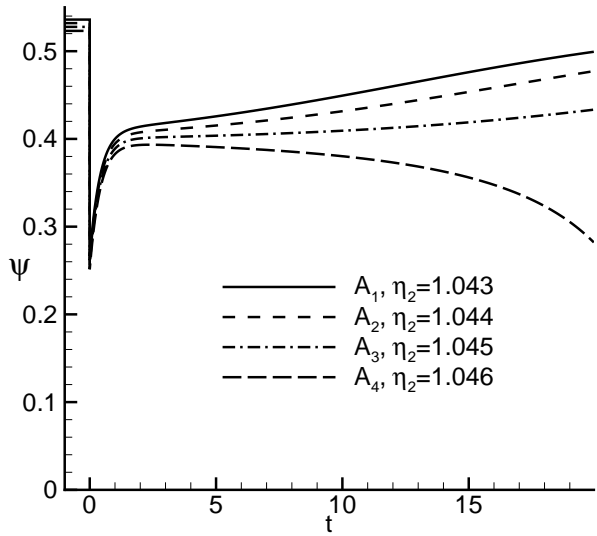,width=0.65\linewidth}  (b)
\end{minipage}\hfill
\caption{\em \small  Values of (a) perturbation growth $J$ and (b)
flow stream function $\Psi$ along the trajectories computed with
Conditional Nonlinear Optimal Perturbation (CNOP), initial
conditions superposed on the thermally-driven state for different
values of $\eta_2$.  } \label{f:10}
\end{figure}

\begin{figure}[htpb]
\begin{minipage}[t]{\linewidth}
\centering\epsfig{file=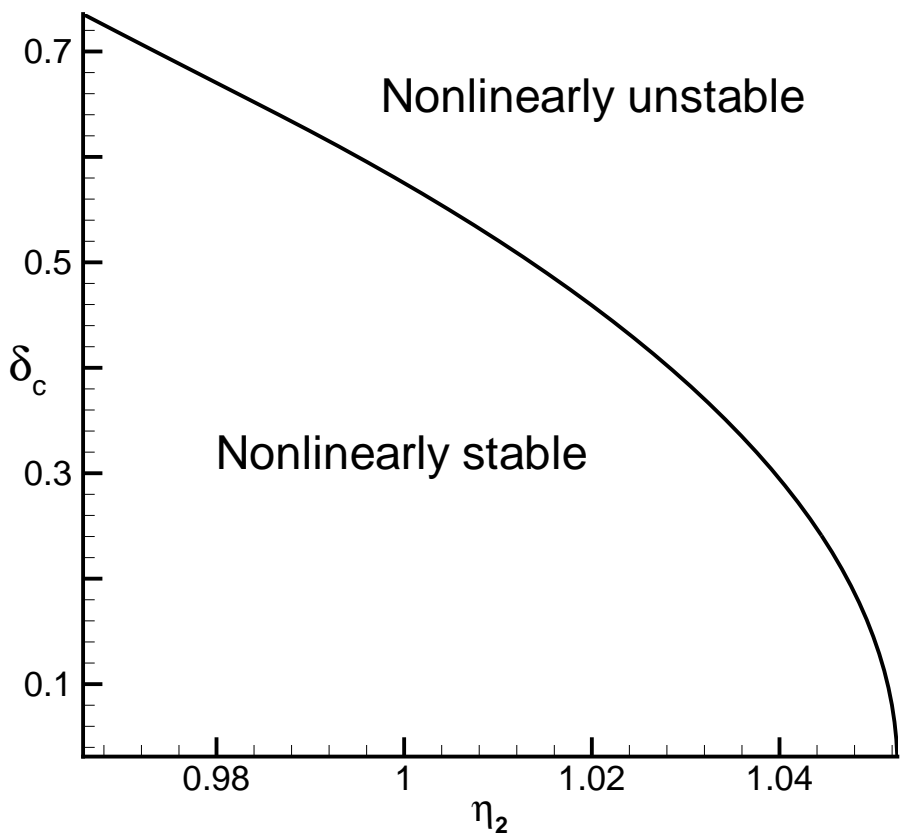,width=\linewidth}
\end{minipage}\hfill
\caption{\em \small   The critical value of $\delta$ ($\delta_c$)
versus the parameter controlling the thermally-driven state near
the saddle-node bifurcation at $\eta_2 = 1.05$. } \label{f:11}
\end{figure}

\begin{figure}[htpb]
\begin{minipage}[t]{\linewidth}
\centering\epsfig{file=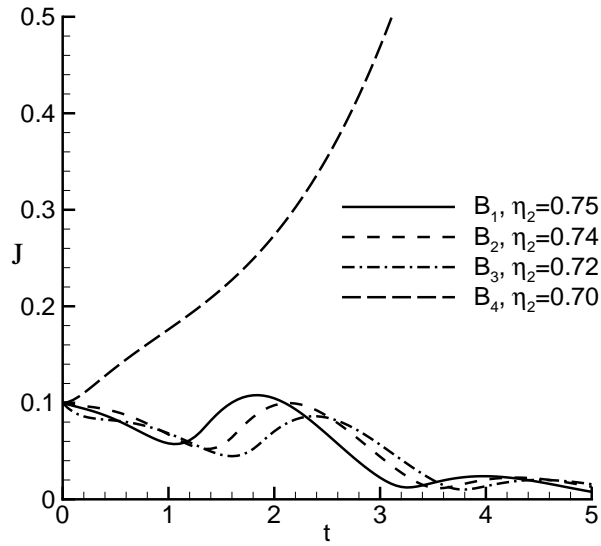,width=0.65\linewidth}  (a) \\
\end{minipage}\hfill
\begin{minipage}[t]{\linewidth}
\centering\epsfig{file=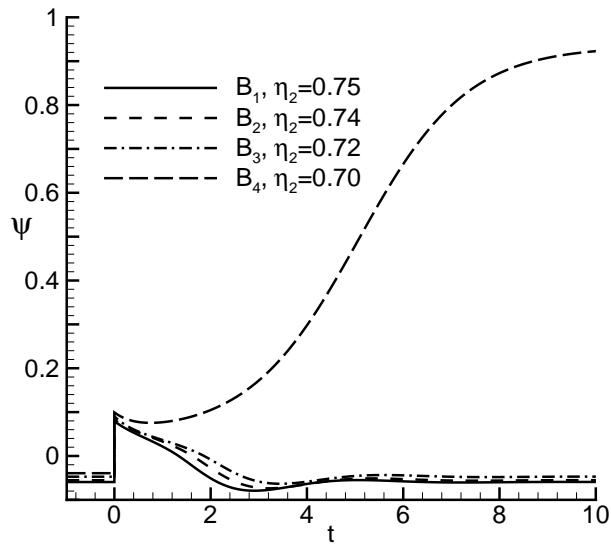,width=0.65\linewidth}  (b)
\end{minipage}\hfill
\caption{\em \small Values of (a) $J$ and (b) $\Psi$ along the
trajectories computed with Conditional Nonlinear Optimal
Perturbation (CNOP) initial conditions superposed on the
salinity-driven state for different values of $\eta_2$.  }
\label{f:12}
\end{figure}

\begin{figure}[htpb]
\begin{minipage}[t]{\linewidth}
\centering\epsfig{file=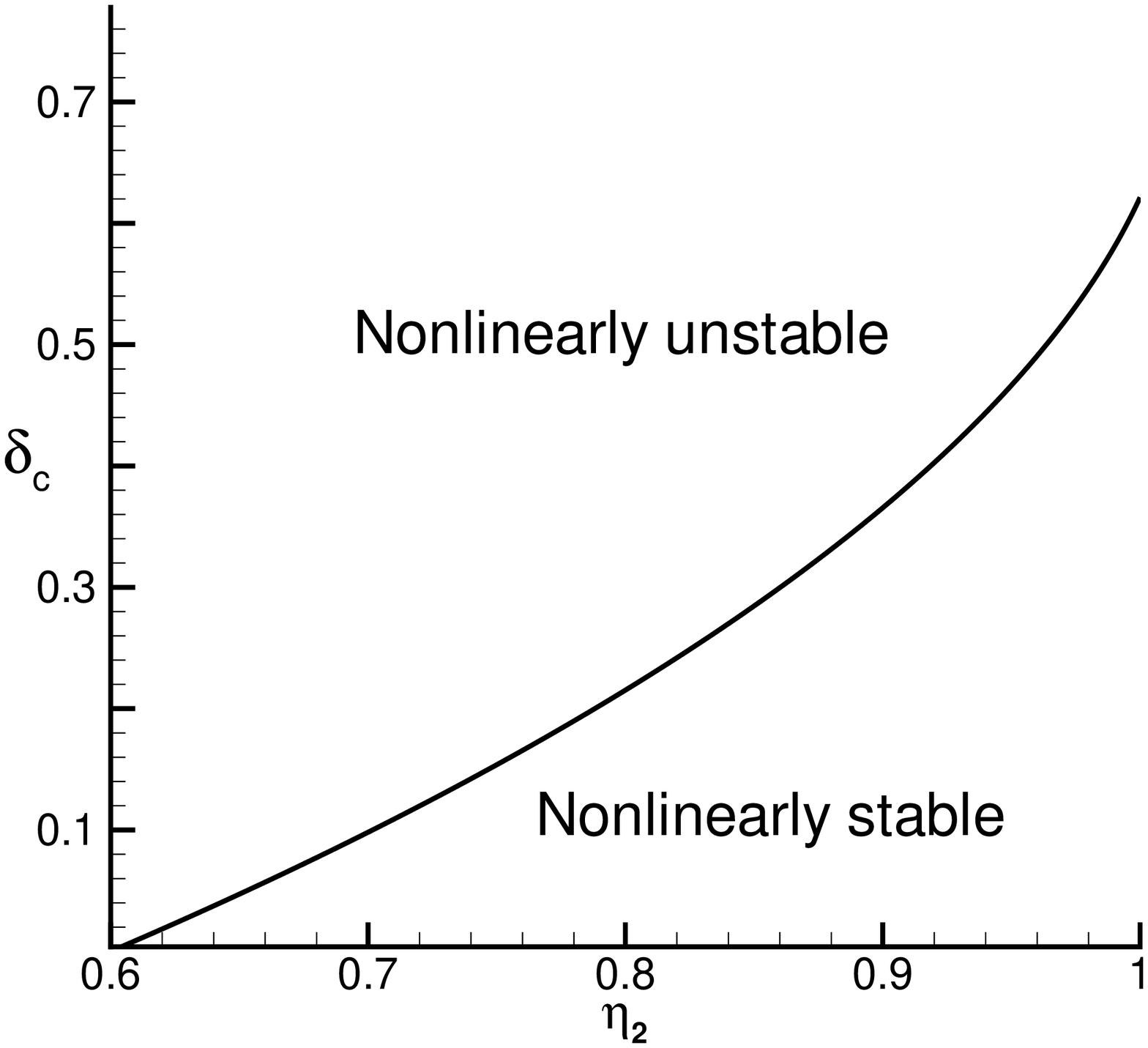,width=\linewidth}
\end{minipage}\hfill
\caption{\em \small The critical value of $\delta$ ($\delta_c$)
versus the parameter controlling the salinity-driven state near
the saddle-node bifurcation at $\eta_2 = 0.6$. } \label{f:13}
\end{figure}

When we use LSV analysis, there are two singular vectors $x_1$ and
$-x_1$ that correspond to one singular value $\sigma_1$ (see
Fig.~\ref{f:4} and Fig.~\ref{f:8}a). If  $x_1$ is a freshwater
type perturbation (\,$\Psi'<0$\,), $-x_1$ must be a salinity type
perturbation (\,$\Psi'>0$\,). The conclusion from the linear
analysis (using LSV) is that the thermohaline circulation is
equally sensitive to either freshwater or salinity entering the
northern North Atlantic. However, the nonlinear analysis (using
CNOP) clearly reveals a difference in  the response of the system
to  the two types of perturbations.

The asymmetry can be understood by considering  the nonlinear
evolution of perturbations in the two-box model. For the TH state,
according to  (\ref{e:thpetur1}), the flow rate
perturbation $\Psi'=T'-S'$ satisfies
\be \frac{d\Psi'}{dt}
=(2\bar{S}-2\bar{T}-1)T'+(2\bar{T}-2\bar{S}+\eta_3)S'-\Psi'^{2}
\label{e:thPsi1} \ee
Integrating the above equation, we find
\be
\Psi'(t)=\Psi'(0)+\int_{0}^{t}L(T',S')d\tau-\int_{0}^{t}\Psi'^{2}d\tau
\label{e:thPsi2} \ee
where $L(T',S')$ is the linear part of (\ref{e:thPsi1}). It is
well known that the two linear terms in (\ref{e:thPsi1}) determine
the linear stability of the steady state \cite[]{Stommel1961}.

For an initial perturbation $\Psi'(0) < 0$ (freshwater type), the
nonlinear term is always negative and  the freshwater perturbation
is amplified. This is a positive feedback and the stronger the
freshwater perturbation, the stronger the nonlinear feedback
destabilizing the TH steady state. A perturbation $\Psi'(0)> 0$
(salinity type) is damped by the negative definite nonlinear term.
This is a negative feedback and  the stronger the salinity
perturbation, the stronger the nonlinear feedback stabilizing the
TH state.  Nonlinear mechanisms hence  make the TH steady  state
more stable to  salinity perturbations.

This explains the results in Fig.~\ref{f:4}. For the freshwater
type  initial perturbations ($\Psi' < 0$), the nonlinear evolution
of the initial perturbation of (\ref{e:thpetur1}) is larger
than the linear evolution of the initial perturbation of
(\ref{e:thpetur2}). For the salinity type initial perturbations
($\Psi' > 0$), the nonlinear evolution is smaller than the linear
 evolution. In general, the nonlinear term yields positive
(negative) feedback for negative (positive) $\Psi'$ in the case of
thermally-dominated (TH) steady states.

On the other hand, when the basic steady flow is a SA state
($\bar{\Psi} < 0$), then we have
\be \frac{d\Psi'}{dt} =(2\bar{T} - 2\bar{S}-1)T'+
(2\bar{S}-2\bar{T}+\eta_3)S' + \Psi'^{2} \ee

Similarly, we have,
\be
\Psi'(t)=\Psi'(0)+\int_{0}^{t}L(T',S')d\tau+\int_{0}^{t}\Psi'^{2}d\tau
 \ee
Hence, due to nonlinear effects the SA steady state becomes more
unstable (stable)  to   disturbances $\Psi' > 0$ ($\Psi' < 0$)
which explains the results in Fig.~\ref{f:8}a. In general, the
nonlinear term yields positive (negative) feedback for positive
(negative) $\Psi'$ in the case of salinity-dominated (SA) steady
states.

The physical mechanism behind the loss of stability of the TH
state is often discussed in terms of the salt-advection feedback
\cite[]{Marotzke1995r}.  A freshwater (salt) perturbation in the
northern North Atlantic decreases (increases) the northward
circulation and hence decreases (increases)  the northward salt
transport.  The salt-advection   feedback is  independent of the
sign of the perturbation  of the  flow rate $\Psi'$.
In contrast, the  nonlinear instability  mechanism of the thermohaline
circulation depends on the sign of the perturbation  of the  flow rate
$\Psi'$ as discussed above.

The CNOP approach also allows us to determine the critical values
of the finite amplitude perturbations (i.e. $\delta_c$) at which
the nonlinearly induced transitions can occur. The techniques are
currently being generalized to be able to apply them to models of
the thermohaline circulation with more degrees of freedom. The aim
is to tackle these problems eventually in global  ocean
circulation models. When applied to the latter  models, the
approach may provide quantitative bounds on perturbations of  the
present thermohaline flow such  that  nonlinear instability can
occur.

\vfill \noindent {\it Acknowledgments:} \ This work was supported
by the National Natural Scientific Foundation of China (Nos.
40233029 and 40221503), and KZCX2-208 of the Chinese Academy of
Sciences. It was initiated during a visit of HD to Beijing in the
Summer of 2002 which was  partially supported from a PIONIER grant
from the Netherlands Organization of Scientific Research (N.W.O.).
We also appreciate the valuable suggestions by anonymous
reviewers.

\vspace{0.5cm}

\centerline{\bf \LARGE Appendix: The SQP  method}

The constrained nonlinear optimization problem considered in this
paper, after discretization and proper transformation of the
objective function $F$,  can be written in the form
$$\min_{x\in R^n} F(x),$$
subject to
$$c_i(x)\leq 0, \ \ \ \mathrm{for} \ i=1, 2, 3, \cdots, n,$$
where the  $c_i$ are constraint functions.  It is assumed that
first derivatives of the problem are known explicitly, i.e., at
any point $x$,  it is possible to compute the gradient
$\bigtriangledown F(x)$ of  $F$ and the Jacobian
$J(x)=\frac{\partial (c_1, c_2, \cdots, c_n)}{\partial (x_1, x_2,
\cdots, x_n)}$ of the constraint functions $c_i$.

The SQP method is an iterative method which solves a quadratic
programming (QP) subproblem at each iteration and it involves
outer and inner iterations. The outer iterations generate a
sequence of iterates $(x^k, \lambda^k)$ that converges to $(x^*,
\lambda^*)$, where $x^*$ and $\lambda^*$ are respectively a
constrained minimizer and the corresponding Lagrange multipliers.
At each iterate,  a QP subproblem is used to generate a search
direction towards the next iterate $(x^{k+1}, \lambda^{k+1})$.
Solving such a subproblem is itself an iterative procedure, which
is therefore regarded as a inner iterate of a SQP method. The
following is an outline of the SQP algorithm used  in this paper.

Step 1. Given a starting iterate $(x^0, \lambda^0)$ and an initial
approximate Hessian $H^0$, set $k=0$.

Step 2. Minor iteration. Solve $d_k$ by the following QP
subproblem.
$$\min_d([\nabla
F(x^k)]^{\top}d^k+\frac{1}{2}(d^{k\top}H^kd^k)),$$ subject to
$$c(x^k)+[\nabla c(x^k)]^{\top}d^k \leq 0,$$
where $d^k$ is a direction of descent for the objective function.

Step 3. Check if $(x^k, \lambda^{k})$ satisfies the convergence
criterion, if not set $x^{k+1}=x^k+\alpha d^k$, where $\alpha\leq
1$. For $\lambda^{k+1}$, it is also determined by $d^k$, which is
automatically realized in the solver  \cite[]{Barclay1997}. Go to Step 4.

Step 4. Update the Hessian Lagrangian by using the BFGS quasi-Newton
formula \cite[]{Liu1989}. Let $s^k=x^{k+1}-x^k$, and $y^k=\nabla
L(x^{k+1}, \lambda^{k+1})-\nabla L(x^k, \lambda^k)$, where $\nabla
L=\nabla F +\nabla c\lambda$. The new Hessian Lagrangian,
$H^{k+1}$, can be obtained by calculating
$$H^{k+1}=H^k-\frac{H^ks^ks^{k\top}H^{k\top}}{s^{k\top}H^ks^k}+
\frac{y^ky^{k\top}}{y^{k\top}s^k}.$$Then set $k=k+1$ and go to
Step 1.

In the SQP algorithm, the definition of the QP Hessian Lagrangian
$H^k$ is crucial to the success of an SQP solver. In
\cite{Gill1997}, $H^k$ is a limited-memory quasi-Newton
approximation to $G= \bigtriangledown^2 L$, the Hessian of the
modified Lagrangian. Another possibility is to define $H^k$ as a
positive-definite matrix related to a finite-difference
approximation to $G$ \cite[]{Barclay1997}. In this paper, we adopt
the former one, which has been shown to be efficient  for  the
nonlinearly constraint optimization problem \cite[]{Mu2003a}.

\newpage
\centerline{\bf Captions to the Figures}
\noindent Fig. 1  \\
The bifurcation diagram of the Stommel two-box model for $\eta_1 =
3.0$ and $\eta_3 = 0.2$.  The points labelled A and B represent
the thermally-driven steady state  and salinity-driven steady
state, respectively, considered in section 3.  The points labelled
P and Q represent the bifurcation points of the model,
respectively. Solid curves indicate linearly stable steady states,
whereas the states on the dashed curve are unstable. There are
thermally-driven (TH) stable steady states ($\bar{\Psi} > 0$) and
salinity-driven (SA, ie., the circulation is salinity-dominated)
stable steady states ($\bar{\Psi}<0$).\\
\medskip
\noindent Fig. 2  \\
The values of $\theta$ for both the linear singular vectors (LSV,
dashed line) and for the Conditional Nonlinear Optimal
Perturbation (CNOP, solid line) of the thermally-driven state
under the conditions $\delta \leq 0.3$ and $t_e = 2.5$. \\
\medskip
\noindent Fig. 3  \\
 Values of $J$ at  the endpoints  of
trajectories  at time $t_e=2.5$ for different values of $\delta$.
These trajectories started either with Conditional Nonlinear
Optimal Perturbation (CNOP-N, solid curve) and linear singular
vectors (LSV-N,dash-dotted curve, hardly distinguishable from the
solid curve) perturbing the thermally-driven state. Also included
are the endpoints when the tangent linear model is integrated with
the linear singular vectors as initial perturbation (LSV-L, dashed
curve ). \\
\medskip
\noindent Fig. 4  \\
The magnitude of perturbations $J$ obtained at $t_e = 2.5$ for the
the evolutions of perturbations of the thermally-driven state in
the tangent linear model and nonlinear model.  The initial
perturbations have the form $(T'(0), S'(0)) =
(\delta\cos\theta,\delta\sin\theta)$ with  $\delta=0.2$. \\
\medskip
\noindent Fig. 5  \\
Values of (a) perturbation growth $J/\delta$ and (b) flow stream
function $\Psi$ along the trajectories computed with Conditional
Nonlinear Optimal Perturbation (CNOP, solid curve) initial
conditions superposed on the thermally-driven
state. The different curves are for different values of $\delta$.   \\
\medskip
\noindent Fig. 6  \\
The values of $\theta$ for both the linear singular vectors (LSV,
dashed line) and for the Conditional Nonlinear Optimal
Perturbations (CNOP) of the salinity-driven
state under the conditions $\delta \leq 0.22$ and $t_e = 2.5$ . \\
\medskip
\noindent Fig. 7  \\
The magnitude of perturbation $J$ at  the endpoints of
trajectories at time $t_e=2.5$ for different values of $\delta$.
These trajectories started either with Conditional Nonlinear
Optimal Perturbation (CNOP-N, solid curve) and linear singular
vectors (LSV-N, dash-dotted line) perturbing the salinity-driven
state. Also included are the endpoints when the tangent linear
model is integrated with the linear singular
vectors as initial perturbation (LSV-L, dashed line).  \\
\medskip
\noindent Fig. 8  \\
(a) Values of perturbation magnitude $J$ obtained at $t_e = 2.5$
for the the evolutions of perturbations of the salinity-driven
state in the tangent linear model and nonlinear model. The initial
perturbations have the form $(T'(0), S'(0)) =
(\delta\cos\theta,\delta\sin\theta)$ with  $\delta=0.2$ .
 And (b) the contour
plot of $J(\theta,\delta)$, with contour interval of 0.02 from
$J=0.02$ to $J=0.08$ (solid curve and dotted curve). The three
group local maxima are indicated as the  dashed curves . \\
\medskip
\noindent Fig. 9  \\
  Values of (a) perturbation growth $J/\delta$
and (b) flow stream function $\Psi$ along the trajectories
computed with Conditional Nonlinear Optimal Perturbation (CNOP),
initial conditions superposed on the salinity-driven state. The
different curves are for different values of $\delta$.  \\
\medskip
\noindent Fig. 10  \\
 Values of (a) perturbation growth $J$ and (b)
flow stream function $\Psi$ along the trajectories computed with
Conditional Nonlinear Optimal Perturbation (CNOP), initial
conditions superposed on the thermally-driven state for different
values of $\eta_2$.  \\
\medskip
\noindent Fig. 11  \\
The critical value of $\delta$ ($\delta_c$) versus the parameter
controlling the thermally-driven state near the saddle-node
bifurcation at $\eta_2 = 1.05$.  \\
\medskip
\noindent Fig. 12  \\
Values of (a) $J$ and (b) $\Psi$ along the trajectories computed
with Conditional Nonlinear Optimal Perturbation (CNOP) initial
conditions superposed on the
salinity-driven state for different values of $\eta_2$. \\
\medskip
\noindent Fig. 13  \\
The critical value of $\delta$ ($\delta_c$) versus the parameter
controlling the salinity-driven state near
the saddle-node bifurcation at $\eta_2 = 0.6$.  \\
\newpage

\newpage

\newpage

\newpage

\newpage

\newpage

\newpage

\newpage

\newpage

\newpage


\begin{thebibliography}{}

\bibitem[\protect\citeauthoryear{Barclay, Gill \& Rosen}{Barclay
  et~al.}{1997}]{Barclay1997}
Barclay, A., P.~E. Gill, and J.~B. Rosen, 1997:
\newblock {SQP} methods and their application to numerical optimal control.
\newblock Technical Report Report NA 97-3, Dept of Mathematics, UCSD.

\bibitem[\protect\citeauthoryear{Berloff \& Meacham}{Berloff and
  Meacham}{1996}]{Barkmeijer1996}
Berloff, P.~S. and S.~P. Meacham, 1996:
\newblock Constructing fast-growing perturbations for the nonlinear regime.
\newblock {\em J.\ Atmos.\ Sci.}, {\bf 53}, 2838--2851.

\bibitem[\protect\citeauthoryear{Cessi}{Cessi}{1994}]{Cessi1994}
Cessi, P., 1994:
\newblock A simple box model of stochastically forced thermohaline flow.
\newblock {\em J.\ Phys.\ Oceanogr.}, {\bf 24}, 1911--1920.

\bibitem[\protect\citeauthoryear{Chen, Battisti, Palmer, Barsugli \&
  Sarachik}{Chen et~al.}{1997}]{ChenYQ1997}
Chen, Y.~Q., S.~D. Battisti, T.~N. Palmer, J.~Barsugli, and E.~S.
Sarachik,
  1997:
\newblock A study of the predictability of tropical pacific sst in a coupled
  atmosphere-ocean model using singular vector analysis: The role of the annual
  cycle and the enso cycle.
\newblock {\em Monthly Weather Review}, {\bf 125}, 831--845.

\bibitem[\protect\citeauthoryear{Dijkstra}{Dijkstra}{2000}]{DijkstraB2000}
Dijkstra, H.~A., 2000:
\newblock {\em {Nonlinear Physical Oceanography: A Dynamical Systems Approach
  to the Large Scale Ocean Circulation and El Ni\~no.}}
\newblock Kluwer Academic Publishers, Dordrecht, the Netherlands.

\bibitem[\protect\citeauthoryear{Durbiano}{Durbiano}{2001}]{Durbiano2001}
Durbiano, S., 2001:
\newblock {\em {Vecteurs caracteristiques de modeles oceaniques pour la
  reduction d'ordre er assimilation de donn\'ees}}.
\newblock PhD thesis, Universit\'e Joseph Fourier, Grenoble, France.

\bibitem[\protect\citeauthoryear{Gill \& Saunders}{Gill and
  Saunders}{1997}]{Gill1997}
Gill, P. E.~Murray, W. and M.~A. Saunders, 1997:
\newblock {\textsc{SNOPT}}: An {\textsc{sqp}} algorithm for large-scale
  constrained optimization.
\newblock Technical Report Report NA 97-2, Dept of Mathematics, UCSD.

\bibitem[\protect\citeauthoryear{Griffies \& Tziperman}{Griffies and
  Tziperman}{1995}]{Griffies1995}
Griffies, S.~M. and E.~Tziperman, 1995:
\newblock A linear thermohaline oscillator driven by stochastic atmospheric
  forcing.
\newblock {\em J. Climate}, {\bf 8}, 2440--2453.

\bibitem[\protect\citeauthoryear{Hasselmann}{Hasselmann}{1976}]{Hasselmann1976}
Hasselmann, K., 1976:
\newblock Stochastic climate models. {I}: {T}heory.
\newblock {\em Tellus}, {\bf 28}, 473--485.

\bibitem[\protect\citeauthoryear{Knutti \& Stocker}{Knutti and
  Stocker}{2002}]{Knutti2002}
Knutti, R. and T.~F. Stocker, 2002:
\newblock Limited predicability of future thermohaline circulation to an
  instability threshold.
\newblock {\em J. Climate}, {\bf 15}, 179--186.

\bibitem[\protect\citeauthoryear{Liu \& Nocedal}{Liu and
  Nocedal}{1989}]{Liu1989}
Liu, D.~C. and J.~Nocedal, 1989:
\newblock on the limited memory method for large scale optimization.
\newblock {\em Mathematical Programming}, {\bf 45}, 503--528.

\bibitem[\protect\citeauthoryear{Manabe \& Stouffer}{Manabe and
  Stouffer}{1999}]{Manabe1999}
Manabe, S. and R.~J. Stouffer, 1999:
\newblock {Are two modes of thermohaline circulation stable?}
\newblock {\em Tellus}, {\bf 51A}, 400--411.

\bibitem[\protect\citeauthoryear{Marotzke}{Marotzke}{1995}]{Marotzke1995r}
Marotzke, J., 1995:
\newblock Analysis of thermohaline feedbacks.
\newblock Technical Report No.39, Center for Global Change Science, M.I.T.
  Cambridge.

\bibitem[\protect\citeauthoryear{McAvaney}{McAvaney}{2001}]{TAR2001}
McAvaney, B., 2001:
\newblock Chapter 8: Model evaluation.
\newblock In {\em {Climate Change 2001: The Scientific Basis}}, Houghton,
  J.~T., Ding, Y., Griggs, D.~J., Noguer, M., van~der Linden, P.~J., and
  Xiaosu, D., editors. Cambridge University Press, 225-256.

\bibitem[\protect\citeauthoryear{Mu, Duan \& Wang}{Mu et~al.}{2003}]{Mu2003b}
Mu, M., W.~Duan, and B.~Wang, 2003:
\newblock {Conditional nonlinear optimal perturbation and its applications}.
\newblock {\em Nonlinear Proc. Geophysics}, {\bf 10}, 493--501.

\bibitem[\protect\citeauthoryear{Mu \& Duan}{Mu and Duan}{2003}]{Mu2003a}
Mu, M. and W.~Duan, 2003:
\newblock {A new approach to study ENSO predictability: conditional nonlinear
  optimal perturbation}.
\newblock {\em Chinese Science Bulletin}, {\bf 48}, 1045--1047.

\bibitem[\protect\citeauthoryear{Mu \& Wang}{Mu and Wang}{2001}]{Mu2001}
Mu, M. and J.~Wang, 2001:
\newblock {Nonlinear fastest growing perturbation and the first kind of
  predictability}.
\newblock {\em Science in China}, {\bf 44D}, 1128--1139.

\bibitem[\protect\citeauthoryear{Mu}{Mu}{2000}]{Mu2000}
Mu, M., 2000:
\newblock {Nonlinear singular vectors and nonlinear singular values}.
\newblock {\em Science in China}, {\bf 43D}, 375--385.

\bibitem[\protect\citeauthoryear{Palmer}{Palmer}{1995}]{Palmer2001}
Palmer, T.~N., 1995:
\newblock {A nonlinear dynamical perspective on model error: A proposal for
  nonlocal stochastic-dynamic parameterisation in weather and climate
  prediction models}.
\newblock {\em Quart. J. Roy. Meteor. Soc.}, {\bf 127}, 279--304.

\bibitem[\protect\citeauthoryear{Rahmstorf}{Rahmstorf}{1995}]{Rahmstorf1995b}
Rahmstorf, S., 1995:
\newblock {Bifurcations of the Atlantic thermohaline circulation in response to
  changes in the hydrological cycle}.
\newblock {\em Nature}, {\bf 378}, 145--149.

\bibitem[\protect\citeauthoryear{Stocker, Wright \& Mysak}{Stocker
  et~al.}{1992}]{Stocker1992}
Stocker, T.~F., D.~G. Wright, and L.~A. Mysak, 1992:
\newblock A zonally averaged, coupled ocean-atmosphere model for paleoclimate
  studies.
\newblock {\em J. Climate}, {\bf 5}, 773--797.

\bibitem[\protect\citeauthoryear{Stommel}{Stommel}{1961}]{Stommel1961}
Stommel, H., 1961:
\newblock Thermohaline convection with two stable regimes of flow.
\newblock {\em Tellus}, {\bf 2}, 244--230.

\bibitem[\protect\citeauthoryear{Thompson}{Thompson}{1998}]{Thompson1998}
Thompson, C.~J., 1998:
\newblock Initial conditions for optimal growth in couple oceanatmosphere model
  of {ENSO}.
\newblock {\em J.\ Atmos.\ Sci.}, {\bf 55}, 537--557.

\bibitem[\protect\citeauthoryear{Timmermann \& Lohmann}{Timmermann and
  Lohmann}{2000}]{Timmermann2000}
Timmermann, A. and G.~Lohmann, 2000:
\newblock Noise-induced transitions in a simplified model of the thermohaline
  circulation.
\newblock {\em J.\ Phys.\ Oceanogr.}, {\bf 30}, 1891--1900.

\bibitem[\protect\citeauthoryear{Tziperman \& Ioannou}{Tziperman and
  Ioannou}{2002}]{Tziperman2002}
Tziperman, E. and P.~J. Ioannou, 2002:
\newblock Transient growth and optimal excitation of thermohaline variability.
\newblock {\em J.\ Phys.\ Oceanogr.}, {\bf 32}, 3427--3435.

\bibitem[\protect\citeauthoryear{V{\'{e}}lez-Belch{\'{i}}, Alvarez, Colet,
  Tintore \& Haney}{V\'{e}lez-Belch\'{i} et~al.}{2001}]{Velez-Belchi2001}
V{\'{e}}lez-Belch{\'{i}}, P., A.~Alvarez, P.~Colet, J.~Tintore,
and R.~L.
  Haney, 2001:
\newblock Stochastic resonance in the thermohaline circulation.
\newblock {\em Geophys.\ Res.\ Letters}, {\bf 28}, 2053--2056.


\bibitem[\protect\citeauthoryear{Xue \& Zebiak}{Xue and
  Zebiak}{1997a}]{Xue1997a}
Xue, Y., M. A.~C. and S.~E. Zebiak, 1997a:
\newblock Predictability of a coupled model of enso using singular vector
  analysis. part i: Optimal growth in seasonal background and enso cycles.
\newblock {\em Monthly Weather Review}, {\bf 125}, 2043¨C2056.

\bibitem[\protect\citeauthoryear{Xue \& Zebiak}{Xue and
  Zebiak}{1997b}]{Xue1997b}
Xue, Y., M. A.~C. and S.~E. Zebiak, 1997b:
\newblock Predictability of a coupled model of enso using singular vector
  analysis. part ii: optimal growth and forecast skill.
\newblock {\em Monthly Weather Review}, {\bf 125}, 2057¨C2073.

\end{thebibliography}
\end{document}